\documentclass[aps,prd,groupedaddress,showpacs,showkeys]{revtex4}
\usepackage{amssymb}
\begin{document}

\vspace*{1cm}  
\begin{center}
{\LARGE\bf Eur.Phys.J.direct C4:18, 2002} 
\end{center}

\vspace*{.5cm} 

\title{The exclusive rare decays $B \to K \bar l l$ and  \\
$B_c \to D(D^\ast) \bar l l$ in a relativistic quark model}
\author{Amand Faessler$^1$, Th.Gutsche$^1$, M.A.Ivanov$^2$, 
J.G.K\"{o}rner$^3$ and V.E.Lyubovitskij$^1$}

\affiliation{$^1$ Institut f\"ur Theoretische Physik, 
Universit\"at T\"ubingen, 
\\ Auf der Morgenstelle 14, D-72076 T\"ubingen, Germany \\ 
$^2$ Bogoliubov Laboratory of Theoretical Physics, 
Joint Institute for Nuclear Research, 141980 Dubna, Russia\\ 
$^3$ Institut f\"{u}r Physik, Johannes Gutenberg-Universit\"{a}t,
D-55099 Mainz, Germany\\}

\begin{abstract}
We study the exclusive rare decay $B\to K \bar l l$.
We calculate the relevant form factors within
a relativistic constituent quark model, for the first time
without employing the impulse approximation. The calculated form
factors are used to evaluate  differential decay rates and
polarization observables. We present results on the $q^2$-dependence 
of a set of observables  with and without long-distance contributions.
A similar analysis is done for the exclusive rare decays
$B_c\to D(D^\ast)\bar l l$ with special emphasis on the cascade decay
$B_c\to D^\ast(\to D\pi)\bar l l$.
We derive a four-fold angular decay distribution for this process
in terms of  helicity amplitudes including lepton mass effects.
The  four-fold angular decay distribution allows to define a number
of physical observables which are amenable to measurement.
We compare our results with the results of other studies. 

\vspace*{.5cm}
\noindent DOI: 10.1007/s1010502c0018 
\end{abstract}


\maketitle

\section{Introduction}
\label{sec:intr}

The flavor-changing neutral current transitions $B \to K + X$ 
and $B_c\to D(D^\ast) + X$ with $X = \gamma, l^+l^-, \bar\nu\nu$ 
are of special interest  because they proceed at the loop level 
in the Standard Model (SM) involving also the top quark. 
They may therefore be used for a determination  of the 
Cabibbo-Kobayashi-Maskawa (CKM) matrix elements $V_{tq}$ $(q=d,s,b)$.
The available experimental measurements of the branching ratio
of the inclusive radiative $B$-meson decay  
$$
\hspace*{-.2cm} 
{\rm Br}\left(B\to X_s\gamma\right)=
\left\{
\begin{array}{lr}
\left(3.11\pm 0.80 ({\rm stat}) \pm 0.72 ({\rm syst})\right)\times 
10^{-4} & 
\mbox{ALEPH~\cite{Barate:1998vz}} \\ 
& \\
\left(3.36\pm 0.53 ({\rm stat})\pm 0.42 ({\rm syst})^{+0.50}_{-0.54}
({\rm th})\right)\times 10^{-4}  & 
\mbox{BELLE~\cite{Abe:2001hk}} \\ 
& \\ 
\left(3.21\pm 0.43 ({\rm stat})\pm 0.27 ({\rm syst})^{+0.18}_{-0.10}
({\rm th})\right)\times 10^{-4}  & 
\mbox{CLEO~\cite{Chen:2001fj}} \\
\end{array}
\right.
$$ 
are consistent with the next-to-leading order prediction of the 
standard model (see, e.g.~\cite{Ali:2001jg} and references therein):   
\begin{equation}
{\rm Br}(B\to X_s\gamma)_{\rm SM}=(3.35\pm 0.30)\times 10^{-4}\,. 
\end{equation}  
The decay  $B\to K\,l^+l^-$ $(l=e,\mu)$ has been 
observed by the BELLE Collaboration \cite{Abe:2001dh} 
with a branching ratio of 
\begin{equation}
\label{bkll}
{\rm Br}\left(B\to K\,l^+l^- \right)=
(0.75^{+0.25}_{-0.21}\pm 0.09)\times 10^{-6}\,.
\end{equation}  
The recent observation of the bottom-charm $B_c$ meson by the 
CDF Collaboration at Tevatron in Fermilab \cite{Abe:1998bc} 
raises hopes that one may  also explore the rare decays of 
the bottom-charm meson in the future.  

The theoretical study of the exclusive rare decays proceeds
in two steps. First, the effective Hamiltonian  for such
transitions is derived by calculating the leading and next-to-leading
loop diagrams in the SM and by using the operator product expansion
and renormalization group techniques. The modern status of this part
of the calculation is described  in the review \cite{Buras:1995} 
(and references therein).     
Second, one needs to evaluate the matrix elements of the 
effective Hamiltonian between  hadronic states. This part of 
the calculation is model dependent since it involves  nonperturbative 
QCD. There are many papers  on this subject. The decay rates, dilepton 
invariant mass spectra and the forward-backforward asymmetry in the 
decays $B\to K\,l^+l^-\,(l=e,\mu,\tau)$ 
have been investigated in the SM and its supersymmetric  
extensions by using improved form factors from light-cone QCD sum rules
\cite{Ali:1999mm}. An updated analysis of  these decays has been
done in \cite{Ali:2001jg} by including  explicit $O(\alpha_s)$
and $\Lambda_{\rm QCD}/m_b$ corrections.
The invariant dilepton mass spectrum and the Dalitz plot
for the  decay $B\to K\,l^+l^-$ have been studied in 
\cite{Greub:1994pi} by using  quark model form factors.
The $B\to K\,l^+l^-$ decay form factors were studied via
QCD sum rules in \cite{Colangelo:1995jv} and  within the 
lattice-constrained
dispersion quark model in \cite{Melikhov:1997wp}. Various aspects
of these decays were discussed in numerous papers by
Aliev {\it et al.}~\cite{Aliev}. 
The exclusive semileptonic rare decays $B\to K\,l^+l^-$
were analyzed in supersymmetric theories in~\cite{Yan:2000dc}.
The angular distribution and CP asymmetries in the decays
$B\to K\pi e^+e^-$ were investigated in~\cite{Kruger:1999xa}.
The lepton polarization for the inclusive decay $B\to X_s l^+l^-$
was discussed in~\cite{Kruger:1996cv} and~\cite{Hewett:1995dk}.
The rare decays of $B_c\to D(D^\ast)\,l^+l^-$ were studied 
in~\cite{Geng:2001vy} by using the form factors evaluated in 
the light front and constituent quark models.
 
In this paper we study the exclusive rare decays $B\to K \bar l l$.  
We employ a relativistic quark model \cite{Ivanov:1996pz,Ivanov:1999ic}
to  calculate the decay  form factors. This model is based on 
an effective Lagrangian which describes the coupling of hadrons $H$ to 
their constituent quarks. The coupling strength  is determined by the 
compositeness condition $Z_H=0$ \cite{SWH,Efimov:zg} where $Z_H$ is the 
wave function renormalization constant of the hadron $H$. 
One starts with an effective Lagrangian written down in
terms of quark and hadron fields. Then, by using  Feynman
rules, the S-matrix elements describing the hadronic interactions are
given in terms of a set of quark diagrams. In particular, the
compositeness condition enables one to avoid a double counting of
hadronic degrees of freedom. The approach is self-consistent and
universally applicable. All calculations of physical observables 
are straightforward. The model has only a small set of 
adjustable  parameters given by  the values of the constituent quark
masses and the scale parameters that define the size of the distribution 
of the constituent quarks inside a given hadron. 
The values of the fit parameters are within the window of expectations.

The shape of the  vertex functions and the  quark  propagators can in
principle  be found from an analysis of the Bethe-Salpeter and
Dyson-Schwinger equations  as was done e.g. in \cite{Ivanov:1998ms}.
In this paper, however, we choose a  phenomenological approach
where the vertex functions are modelled by a Gaussian form, the size
parameter of which is determined by a fit to the leptonic and radiative
decays of the lowest lying charm and bottom mesons. For the quark 
propagators we use the local representation. In the present calculations 
we do not employ the so-called impulse approximation used 
previously \cite{Ivanov:1999ic}. 
The numerical results obtained with and without the impulse approximation
are close to each other for light-to-light and heavy-to-heavy transitions
but differ considerably from one another for  heavy-to-light transitions 
as e.g. in the  $B\to \pi$ transitions.  

We calculate the form factors of the transition $B\to K$ and use them 
to evaluate differential decay rates and polarization observables. 
We give the $q^2$-dependence of a set of observables with and without  
long-distance contributions which include the lower-lying charmonium 
states according to \cite{Ali:1991is}. We extend our analysis to the 
exclusive rare decay $B_c\to D(D^\ast)\bar ll$. We derive a four-fold 
angular decay distribution for the cascade 
$B_c\to D^\ast(\to D \pi)\bar ll$ process in the  helicity frame 
including lepton mass effects following the method outlined  
in~\cite{Korner:1989qb}. The four-fold angular decay distribution 
allows one to define  a number of physical observables which are 
amenable to measurement. We compare our results with the ones of other 
studies. 

We should remark that our approach is developed mainly 
for the hadrons (mesons and baryons) which satisfy to 
the so-called "threshold inequality": the hadron mass 
should be smaller the sum of their constituents, i.e. 
the sum of the constituent quark masses. In this vein, 
our model was successfully developed for the study of light 
hadrons (e.g., pion, kaon, baryon octet, $\Delta$-resonance), 
heavy-light hadrons (e.g., $D$, $D_s$, $B$ and $B_s$-mesons, 
$\Lambda_Q$, $\Sigma_Q$, $\Xi_Q$ and $\Omega_Q$-baryons) 
and double heavy hadrons (e.g, $J/\Psi$, $\Upsilon$ and 
$B_c$-mesons, $\Xi_{QQ}$ and $\Omega_{QQ}$ 
baryons)~\cite{Ivanov:1996pz,Ivanov:1999ic}.  
To extend our approach to other hadrons we had to introduce 
extra model parameters or do some approximations, like, e.g., 
to introduce the cutoff parameter for external hadron momenta 
to guarantee the fulfilment of the mentioned above 
"threshold inequality". Therefore, at the present stage 
we can not apply our approach for the study of rare decays 
involving $K^\ast$ mesons. Probably, it will be a subject 
of our future investigations. 

The layout of the paper is as follows. In Sec. II we discuss the
effective Hamiltonian. We use the analytical expressions
for the Wilson coefficients from~\cite{Buras:1995} and
the input parameters from \cite{Ali:1999mm}. 
Sec. III is devoted to the description of the  
$B(B_c)\to K(D, D^\ast) \bar ll$ decays in terms of helicity amplitudes.
We derive the four-fold angular decay distribution for the cascade  
$B_c\to D^\ast(\to D \pi)\bar ll$ process 
in the helicity system and define a number of physical observables 
from the angular decay distribution. 
Our analysis goes beyond the results on the four-fold decay distribution
presented in \cite{Kim:2000dq,Ali:2002qc} in that we include 
lepton mass effects appropriated for the treatment of the 
channel with $\tau$-leptons in the final state. 
We also present results on the longitudinal polarization
of the leptons in the $\bar ll$-CM frame. 
This analysis differs from the analysis in \cite{Kim:2000dq,Ali:2002qc}
where the polarization of the leptons were calculated in the initial 
meson rest system. We also include lepton mass effects in the 
polarization calculation. In Sec. IV we briefly
discuss our relativistic quark model and demonstrate the difference
between the exact calculation and the approximate calculation
using the impulse approximation taking as an example the $B-\pi$ form 
factor. We calculate  matrix elements and form factors for the decay
$B\to K \bar ll$ and compare their behavior 
with those calculated in \cite{Ali:1999mm}.
In Sec. V we present our numerical results for branching ratios and
asymmetry parameters. We plot the $q^2$-dependence of the differential
decay rate and the longitudinal polarization of the leptons 
with and without long distance contributions.   
 
\section{Effective Hamiltonian}
\label{sec:ham}

The starting point of the description of the rare exclusive decays
is the effective Hamiltonian obtained from the SM-diagrams 
by using the operator product expansion and renormalization 
group techniques. It allows one to separate the short-distance 
contributions and isolate  them in the Wilson coefficients which can be 
studied systematically within perturbative QCD. The long-distance
contributions are contained in the matrix elements of local operators.
Contrary to the short-distance contributions the calculation of
such matrix elements requires  nonperturbative methods and is therefore
model dependent.

We will follow Refs.\cite{Buras:1995} in writing down the
analytical expressions for the effective Hamiltonian and
paper \cite{Ali:1999mm} in using the numerical values of
the input parameters characterizing the short-distance contributions. 
At the quark level, the rare semileptonic decay 
$b \to s(d) l^+ l^-$ can be described in terms of the effective 
Hamiltonian:  
\begin{equation}
H_{\rm eff} = - \frac{G_F}{\sqrt{2}} \lambda_t   
              \sum_{i=1}^{10} C_i(\mu)  Q_i(\mu) \; . 
\end{equation}
where $\lambda_t \equiv V_{ts(d)}^\dagger V_{tb}$ is the product 
of CKM elements. For example, the standard set \cite{Buras:1995} 
of local operators for $b \to s l^+ l^-$ transition is written as 
\begin{eqnarray} 
\begin{array}{ll} 
Q_1     =  (\bar{s}_{i}  c_{j})_{V-A} ,
           (\bar{c}_{j}  b_{i})_{V-A}  ,               &
Q_2     =  (\bar{s} c)_{V-A}  (\bar{c} b)_{V-A}   ,              \\
Q_3     =  (\bar{s} b)_{V-A}\sum_q(\bar{q}q)_{V-A}  ,            &
Q_4     =  (\bar{s}_{i}  b_{j})_{V-A} \sum_q (\bar{q}_{j}
          q_{i})_{V-A} ,                                    \\
Q_5     =  (\bar{s} b)_{V-A}\sum_q(\bar{q}q)_{V+A} ,            &
Q_6     =  (\bar{s}_{i}  b_{j })_{V-A} 
   \sum_q  (\bar{q}_{j}  q_{i})_{V+A}  ,               \\
Q_7     =  \frac{e}{8\pi^2} m_b (\bar{s} \sigma^{\mu\nu}
          (1+\gamma_5) b) F_{\mu\nu}  ,                     &
Q_8    =  \frac{g}{8\pi^2} m_b (\bar{s}_i \sigma^{\mu\nu}
   (1+\gamma_5) {\bf T}_{ij} b_j) {\bf G}_{\mu\nu}  ,          \\
Q_9     = \frac{e}{8\pi^2} (\bar{s} b)_{V-A}  (\bar{l}l)_V ,      &
Q_{10}  = \frac{e}{8\pi^2} (\bar{s} b)_{V-A}  (\bar{l}l)_A  
\end{array}
\end{eqnarray}
where ${\bf G}_{\mu\nu}$ and $F_{\mu\nu}$ are the gluon and photon 
field strengths, respectively; ${\bf T}_{ij}$ are the generators of 
the $SU(3)$ color group; $i$ and $j$ denote color indices 
(they are omitted in the color-singlet currents).
Labels $(V\pm A)$ stand for 
$\gamma^\mu(1\pm\gamma^5)$.  $Q_{1,2}$ are current-current operators, 
$Q_{3-6}$ are QCD penguin operators,  $Q_{7,8}$ are "magnetic  
penguin" operators, and $Q_{9,10}$ are semileptonic electroweak penguin 
operators. Explicit formulae for the Wilson coefficients $C_i(\mu)$ 
obtained in leading logarithmic order are written down in Appendix A. 

The effective Hamiltonian leads to the free quark 
$b\to s l^+l^-$-decay amplitude: 
\begin{eqnarray}
M(b\to s\ell^+\ell^-) & = & 
\frac{G_F \alpha}{2\sqrt{2}  \pi} \,\lambda_t \, 
\left\{
 C_9^{\rm eff}\,
       \left( \bar{s}  b \right)_{V-A} \,\left(  \bar l l \right)_V
+C_{10}\left( \bar{s}  b \right)_{V-A} \,\left(  \bar l l \right)_A
\right.
\label{free}\\
&-& \left. \frac{2m_b}{q^2}\,C_7^{\rm eff}\, 
\left( \bar{s}\,i\sigma^{\mu \nu}\,(1+\gamma^5)\, q^\nu \,b \right)\,
                        \left( \bar l l \right)_V \right\}. \nonumber
\end{eqnarray}
where $C_7^{\rm eff}= C_7 -C_5/3 -C_6$. 

The Wilson coefficient $ C_9^{\rm eff}$ effectively takes 
into account, first, the contributions from the four-quark
operators $Q_i$ ($i=1,...,6$) and, second, the nonperturbative 
effects coming from the $c\bar c$-resonance contributions
which are as usual parametrized by a Breit-Wigner ansatz 
\cite{Ali:1991is}:

\begin{eqnarray}
C_9^{\rm eff} & = & C_9 + 
C_0 \left\{
h(\hat m_c,  s)+ \frac{3 \pi}{\alpha^2}\,  \kappa\,
         \sum\limits_{V_i = \psi(1s),\psi(2s)}
      \frac{\Gamma(V_i \rightarrow l^+ l^-)\, m_{V_i}}
{  {m_{V_i}}^2 - q^2  - i m_{V_i} \Gamma_{V_i}}
\right\} 
 \nonumber \\
&-& \frac{1}{2} h(1,  s) \left( 4 C_3 + 4 C_4 +3 C_5 + C_6\right)  
 \\
&-& \frac{1}{2} h(0,  s) \left( C_3 + 3 C_4 \right) +
\frac{2}{9} \left( 3 C_3 + C_4 + 3 C_5 + C_6 \right)\,.
\nonumber
\end{eqnarray}
where $C_0\equiv 3 C_1 + C_2 + 3 C_3 + C_4+ 3 C_5 + C_6$.
Here
\begin{eqnarray*} 
h(\hat m_c,  s) & = & - \frac{8}{9}\ln\frac{m_b}{\mu} 
- \frac{8}{9}\ln\hat m_c +
\frac{8}{27} + \frac{4}{9} x \nonumber\\
& - & \frac{2}{9} (2+x) |1-x|^{1/2} \left\{
\begin{array}{ll}
\left( \ln\left| \frac{\sqrt{1-x} + 1}{\sqrt{1-x} - 1}\right| - i\pi 
\right), &
\mbox{for } x \equiv \frac{4 \hat m_c^2}{ s} < 1 \nonumber \\
 & \\
2 \arctan \frac{1}{\sqrt{x-1}}, & \mbox{for } x \equiv \frac
{4 \hat m_c^2}{ s} > 1,
\end{array}
\right. \nonumber\\
h(0,  s) & = & \frac{8}{27} -\frac{8}{9} \ln\frac{m_b}{\mu} - 
\frac{4}{9} \ln\ s + \frac{4}{9} i\pi.\nonumber 
\end{eqnarray*}
where $\hat m_c=m_c/m_B$, $s=q^2/m_B^2$ and $\kappa=1/C_0$.

The relation between the $\overline{MS}$ b-quark mass $m_b\equiv 
m_b(\mu)$ at a scale $\mu$ and its pole mass $m_{b,\rm pole}$ 
is given by
\begin{eqnarray}
m_b(\mu)=m_{b,\rm pole}\left\{1-\frac{4}{3}\,\frac{\alpha_s(\mu)}{\pi}
\left[1-\frac{3}{4}\ln\biggl(\frac{m_{b,\rm pole}}{\mu^2}\biggr)\right]
\right\}
\end{eqnarray}
where
\begin{eqnarray}
\alpha_s(\mu)=\frac{4\pi}{\beta_0
\ln\frac{\displaystyle{\mu^2}}{\displaystyle{\Lambda^2_{QCD}}}}\,
\left\{1-\frac{\beta_1}{\beta_0^2} 
\frac{\ln\ln\frac{\displaystyle{\mu^2}}{\displaystyle{\Lambda^2_{QCD}}}}
{\ln\frac{\displaystyle{\mu^2}}{\displaystyle{\Lambda^2_{QCD}}}}\right\}
\end{eqnarray}
and where $\beta_0 = 23/3$ and $\beta_1 = 116/3$, as is appropriate 
for five flavors. 
We will use a scale $\mu=m_{b,\rm pole}$ throughout this paper.
The numerical values of the input parameters are 
taken from  \cite{Ali:1999mm} and the corresponding values 
of the Wilson coefficients used in the numerical calculations 
are listed in Table~\ref{tab:input}. 

\section{$B\to K \bar l l$ and $B_c\to D(D^\ast) 
\bar l l$-decays}\label{sec:bk}

\subsection{Form factors and differential decay distributions}

We specify our choice of the momenta as $p_1=p_2+k_1+k_2$ with 
$p_1^2=m_1^2$, $p_2^2=m_2^2$ and  $k_1^2=k_2^2=\mu^2$ where $k_1$ and 
$k_2$ are the $l^+$ and $l^-$ momenta, and $m_1$, $m_2$, $\mu$ are 
the masses of initial meson $H_{in}$, final meson $H_f$ 
and lepton, respectively.

The matrix elements of the exclusive transitions 
$B\to K \bar l l$ and $B_c\to D(D^\ast) \bar l l$ 
are defined by 

\begin{eqnarray}
M(H_{in}\to H_f \bar l l) & = & 
\frac{G_F}{\sqrt{2}}\cdot\frac{ \alpha\lambda_t}{2\,\pi} \cdot 
\left\{C_9^{\rm eff}\,
<H_f\,|\,\bar{s}\,O^\mu\, b\,|\,H_{in}> \,  \bar l\gamma^\mu l
\right.
\label{exclus}\\
&+&
C_{10}\, <H_f\,|\,\bar{s}\,O^\mu\, b\,|\,H_{in}>
\, \bar l\gamma^\mu \gamma^5 l
\nonumber\\
&-& 
\left.
\frac{2m_b}{q^2}\,C_7^{\rm eff}\, 
<H_f\,|\,  \bar{s}\,i\sigma^{\mu \nu}\,(1+\gamma^5)\, 
q^\nu \,b\,|\,H_{in}> \,  \bar l\gamma^\mu l 
\right\}\,.
\nonumber
\end{eqnarray}
where $H_{in} = B$ or $B_c$, $H_f = K, D$ or $D^\ast$. 

We define dimensionless form factors by
\begin{eqnarray}
&&
<K(D)(p_2)\,|\,\bar s(d)\, \gamma_\mu\, b\,| B(B_c)(p_1)>= 
F_+(q^2) P_\mu+F_-(q^2) q_\mu\,,
\label{ff}\\
&&\nonumber\\
&&
<K(D)(p_2)\,|\,\bar s(d) \,i\sigma_{\mu\nu}q^\nu\, b\,|\,B(B_c)(p_1)>=
- \frac{1}{m_1+m_2} \, P_\mu^\perp \, q^2\, F_T(q^2)\,,\nonumber\\
&&\nonumber\\
&&
i<D^\ast(p_2,\epsilon_2)\,|\,\bar d\, O_\mu\, b\,|\,B_c(p_1)> =
\frac{1}{m_1+m_2}\,\epsilon_2^{\dagger\nu}\nonumber\\
& &\times \{
-g_{\mu\nu}\,Pq\,A_0(q^2) +P_\mu P_\nu\,A_+(q^2)
+q_\mu P_\nu\,A_-(q^2)+
i\varepsilon_{\mu\nu\alpha\beta} P^\alpha q^\beta 
\,V(q^2)\}\,,
\nonumber\\
&&\nonumber\\
&&i<D^\ast(p_2,\epsilon_2)\,|\,
\bar d\, i\sigma_{\mu\nu}q^\nu(1+\gamma_5)\, b \,|\,B_c(p_1)>= 
\nonumber\\
& & = \epsilon_2^{\dagger\nu}\{ \, g_{\mu\nu}^\perp \,Pq 
\,a_0(q^2) - P_\mu^\perp \, P_\nu\,a_+(q^2)
-i\varepsilon_{\mu\nu\alpha\beta} P^\alpha q^\beta \,g(q^2)\}
\nonumber
\end{eqnarray}
where $P=p_1+p_2$, $q=p_1-p_2$, \, 
$P_\mu^\perp \doteq P_\mu - q_\mu Pq/q^2$, \,  
$g_{\mu\nu}^\perp \doteq g_{\mu\nu} - q_\mu q_\nu/q^2$, \,  
and $\epsilon^\dagger_2$ is the polarization four-vector 
of the $D^\ast$. Since we want to compare our calculations with 
those in \cite{Ali:1999mm} and \cite{Geng:2001vy}, it is useful
to relate our form factors to those used in \cite{Ali:1999mm} 
and \cite{Geng:2001vy}. The relations read 

\begin{eqnarray}
F_+ &=& f_+^{\rm Ali}=F_+^{\rm Geng}
\nonumber\\
F_- &=& -\,\frac{m_1^2-m_2^2}{q^2}\,(f_+ - f_0)^{\rm Ali}=F_-^{\rm Geng}
\nonumber\\
F_T &=& f_T^{\rm Ali}=-\,F_T^{\rm Geng}
\nonumber\\
&&\nonumber\\
A_0 &=& \frac{m_1 + m_2}{m_1 - m_2}\,A_1^{\rm Ali}=-\,A_0^{\rm Geng}
\nonumber\\
A_+ &=& A_2^{\rm Ali}=A_+^{\rm Geng}
\nonumber\\
A_- &=& \frac{2m_2(m_1+m_2}{q^2}\,(A_3 - A_0)^{\rm Ali}=A_-^{\rm Geng}
\nonumber\\
V &=& V^{\rm Ali}=-\,V^{\rm Geng}
\nonumber\\
&&\nonumber\\
a_0 &=& T_2^{\rm Ali}=\,a_0^{\rm Geng}
\nonumber\\
a_+ &=& (T_2+\frac{q^2}{m_1^2-m_2^2}\,T_3)^{\rm Ali}=-\,a_+^{\rm Geng} 
\nonumber\\
g &=& T_1^{\rm Ali}=g^{\rm Geng}\nonumber
\end{eqnarray}

The matrix element in Eq~(\ref{exclus}) is written as

\begin{equation}
M\left(B(B_c)\to K(D^\ast)\bar ll\right)=
\frac{G_F}{\sqrt{2}}\cdot\frac{\alpha\lambda_t}{2\pi}\,
\left\{
T_1^\mu\,(\bar l\gamma_\mu l)+T_2^\mu\,(\bar l\gamma_\mu\gamma_5 l)
\right\} 
\end{equation}
where the quantities $T_i^\mu$ are expressed through the form
factors and the Wilson coefficients in the following manner:

\noindent
{\bf (a)\,} $B(B_c)\to K(D) \bar ll$-decay:

\begin{eqnarray}
T_i^\mu &=& {\cal F}_+^{(i)}\,P^\mu+ {\cal F}_-^{(i)}\,q^\mu
\hspace{2cm} (i=1,2)\,,
\label{amp_pp}\\
&&\nonumber\\
{\cal F}_+^{(1)} &=& C_9^{\rm eff}\,F_+ + C_7^{\rm eff}\,F_T\, 
\frac{2m_b}{m_1+m_2}\,,
\nonumber\\
{\cal F}_-^{(1)} &=& C_9^{\rm eff}\,F_- - C_7^{\rm eff}\,F_T\,
\frac{2m_b}{m_1+m_2}\,
\frac{Pq}{q^2}\,,
\nonumber\\
&&\nonumber\\
{\cal F}_\pm^{(2)} &=& C_{10}\,F_\pm\,.\nonumber
\end{eqnarray}

\noindent
{\bf (b)\,} $B_c\to D^\ast \bar ll$-decay:

\begin{equation}
\label{amp_pv}
T_i^\mu =  T_i^{\mu\nu}\,\epsilon^\dagger_{2\nu}\,, \hspace{1cm} 
(i=1,2)\,,
\end{equation}

\begin{eqnarray*} 
T_i^{\mu\nu} &=& \frac{1}{m_1+m_2} \{
- Pq\,g^{\mu\nu}\,A_0^{(i)}+P^\mu P^\nu\,A_+^{(i)}+q^\mu P^\nu\, 
A_-^{(i)} +i\varepsilon^{\mu\nu\alpha\beta}P_\alpha q_\beta\,V^{(i)}  
\} \\
&&\\
V^{(1)} &=&   C_9^{\rm eff}\,V + C_7^{\rm eff}\,g\, 
\frac{2m_b(m_1+m_2)}{q^2}\,,\\
A_0^{(1)} &=& C_9^{\rm eff}\,A_0 + C_7^{\rm eff}\,a_0\,
\frac{2m_b(m_1+m_2)}{q^2}\,,\\
A_+^{(1)} &=& C_9^{\rm eff}\,A_+ + C_7^{\rm eff}\,a_+\,
\frac{2m_b(m_1+m_2)}{q^2}\,,\\
A_-^{(1)} &=& C_9^{\rm eff}\,A_- + C_7^{\rm eff}\,(a_0-a_+)\,
\frac{2m_b(m_1+m_2)}{q^2}\,
\frac{Pq}{q^2}\,,\\
V^{(2)}   &=& C_{10}\,V\,, \hspace{1cm}
A_0^{(2)} = C_{10}\,A_0\,,\hspace{1cm}
A_\pm^{(2)} = C_{10}\,A_\pm\,.
\end{eqnarray*}

\begin{figure}[t]
 \vspace*{12cm}
\includegraphics{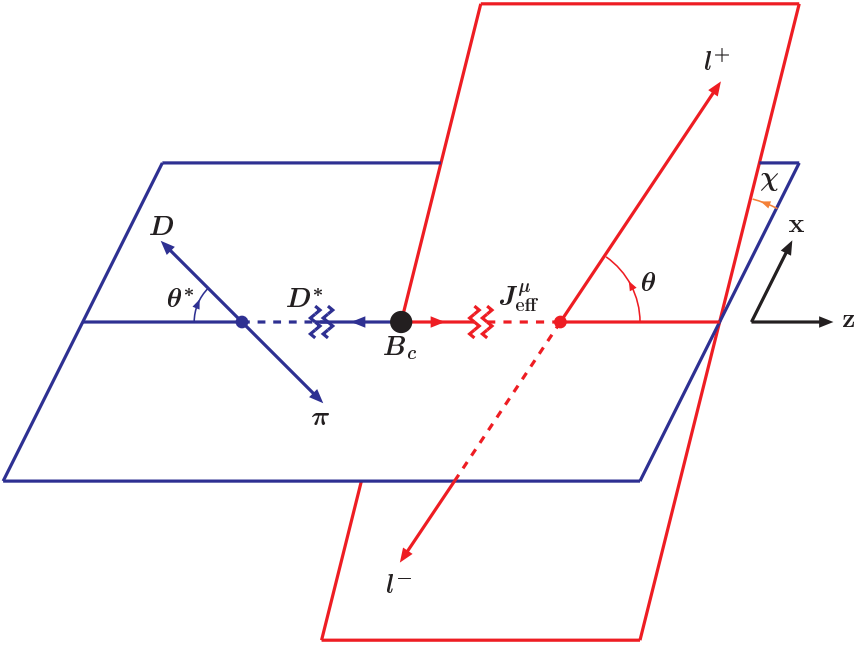}
\vspace*{-1.5cm}
\caption{Definition of angles $\theta$, $\theta^\ast$ and $\chi$ in
the cascade decay $B_c\to D^\ast(\to D\pi)\bar l l$.}
\label{fig:bkangl}
\end{figure}

Let us first consider the polar angle decay distribution differential
in the momentum transfer squared $q^2$. The polar angle is defined
by the angle between $\vec q=\vec p_1-\vec p_2$ and $\vec k_1$
($l^+l^-$ rest frame) as shown in Fig.~\ref{fig:bkangl}. One has

\begin{eqnarray}
\frac{d^2\Gamma}{dq^2 d\cos\theta} &=& 
\frac{|{\bf p_2}| \, v}{(2\pi)^3\,4\,m_1^3}
\cdot\frac{1}{8}\sum\limits_{\rm pol}|M|^2
=\frac{G^2_F}{(2\pi)^3}\, 
\biggl(\frac{\alpha|\lambda_t|}{2\,\pi}\biggr)^2
\frac{|{\bf p_2}|\,v }{8 m_1^2}\label{distr}\\
&\times& \frac{1}{8}
\biggl\{H^{\mu\nu}_{11}\cdot
{\rm tr} [\gamma_\mu\,(\not\! k_1-\mu)\,\gamma_\nu\,
(\not\! k_2+\mu) ]\nonumber\\
&&\nonumber\\
&+&H^{\mu\nu}_{22}\cdot {\rm tr} [\gamma_\mu\gamma_5 \,
(\not\! k_1-\mu)\,\gamma_\nu\gamma_5\,(\not\! k_2+\mu) ]\nonumber\\
&&\nonumber\\
&+&\,H^{\mu\nu}_{12}\cdot {\rm tr} [\gamma_\mu\,(\not\! k_1-\mu)\,
              \gamma_\nu\gamma_5\,(\not\! k_2+\mu) ]\nonumber\\
&&\nonumber\\
&+&H^{\mu\nu}_{21}\cdot
{\rm tr} [\gamma_\mu\gamma_5\,(\not\! k_1-\mu)\,
              \gamma_\nu\,(\not\! k_2+\mu) ]
\biggr\}\nonumber\\
&&\nonumber\\
&=&
\frac{G^2_F}{(2\pi)^3}\,\biggl(\frac{\alpha|\lambda_t|}{2\,\pi}\biggr)^2
\frac{|{\bf p_2}|\,v } {8 m_1^2}
\cdot\frac{1}{2} \biggl\{
 L^{(1)}_{\mu\nu}\cdot (H^{\mu\nu}_{11}+H^{\mu\nu}_{22})
\nonumber\\
&&\nonumber\\
&-& \frac{1}{2}\,L^{(2)}_{\mu\nu}\cdot 
(q^2\,H^{\mu\nu}_{11}+  (q^2-4\mu^2)\,H^{\mu\nu}_{22})
+L^{(3)}_{\mu\nu}\cdot (H^{\mu\nu}_{12}+H^{\mu\nu}_{21})
\biggr\}\nonumber
\end{eqnarray}
where  $|{\bf p_2}|=\lambda^{1/2}(m_1^2,m_2^2,q^2)/2m_1$ is
the momentum of the $K(D^\ast)$-meson and $v=\sqrt{1-4\mu^2/q^2}$ 
is the lepton velocity both given in the $B(B_c)$-rest frame. 
We have introduced lepton and hadron tensors as

\begin{eqnarray}
L^{(1)}_{\mu\nu} &=& k_{1\mu} k_{2\nu}+ k_{2\mu} k_{1\nu}\,,
\hspace{1cm}
L^{(2)}_{\mu\nu}= g_{\mu\nu}\,,
\hspace{1cm}
L^{(3)}_{\mu\nu}= i\varepsilon_{\mu\nu\alpha\beta}k_1^\alpha k_2^\beta\,,
\nonumber\\
&&\label{tensors}\\
H_{ij}^{\mu\nu} & = &T_i^\mu\,T_j^{\dagger\nu}\,.
\nonumber
\end{eqnarray}

\subsection{Helicity amplitudes and two-fold distributions}

The Lorentz contractions in Eq.~(\ref{distr}) can be  evaluated in 
terms of helicity amplitudes as described  in \cite{Korner:1989qb}.
First, we define an orthonormal and complete helicity basis
$\epsilon^\mu(m)$ with the three spin 1 components  orthogonal to
the momentum transfer $q^\mu$, i.e. $\epsilon^\mu(m) q_\mu=0$ for 
$m=\pm,0$, and the spin 0 (time)-component $m=t$ with
$\epsilon^\mu(t)= q^\mu/\sqrt{q^2}$.   

The orthonormality and completeness properties read
 
\begin{eqnarray}
&&
\epsilon^\dagger_\mu(m)\epsilon^\mu(n)=g_{mn} \hspace{1cm}
(m,n=t,\pm,0), 
\nonumber\\
&&\label{orthonorm}\\
&&
\epsilon_\mu(m)\epsilon^{\dagger}_{\nu}(n)g_{mn}=g_{\mu\nu}
\nonumber
\end{eqnarray}
with $g_{mn}={\rm diag}\,(\,+\,,\,\,-\,,\,\,-\,,\,\,-\,)$.
We include the time component polarization vector $\epsilon^\mu(t)$
in the set because we want to discuss lepton mass effects in the 
following.

Using the completeness property we rewrite the contraction
of the lepton and hadron tensors in Eq.~(\ref{distr}) according to
\begin{eqnarray}
L^{(k)\mu\nu}H_{\mu\nu}^{ij} &=& 
L_{\mu'\nu'}^{(k)}\epsilon^{\mu'}(m)\epsilon^{\dagger\mu}(m')g_{mm'}
\epsilon^{\dagger \nu'}(n)\epsilon^{\nu}(n')g_{nn'}H_{\mu\nu}^{ij}
\nonumber\\
&&\nonumber\\
&=& L^{(k)}(m,n)g_{mm'}g_{nn'}H^{ij}(m',n')
\label{contraction}
\end{eqnarray}
where we have introduced the lepton and hadron tensors in the space
of the helicity components

\begin{eqnarray}
\label{hel_tensors}
L^{(k)}(m,n) = \epsilon^\mu(m)\epsilon^{\dagger \nu}(n)
L^{(k)}_{\mu\nu},
\hspace*{.5cm}
H^{ij}(m,n) = \epsilon^{\dagger\mu}(m)\epsilon^\nu(n)H^{ij}_{\mu\nu}\, .
\end{eqnarray}
The point is that the two tensors can be evaluated in two different
Lorentz systems. The lepton tensors $L^{(k)}(m,n)$ will be evaluated
in the $\bar ll$-CM system whereas the hadron tensors $H^{ij}(m,n)$
will be evaluated in the $B(B_c)$ rest system.

In the $B(B_c)$ rest frame one has 

\begin{eqnarray}
p^\mu_1 &=& (\,m_1\,,\,\,0,\,\,0,\,\,0\,)\,,
\nonumber\\
p^\mu_2 &=& (\,E_2\,,\,\,0\,,\,\,0\,,\,\,-|{\bf p_2}|\,)\,,
\\
q^\mu   &=& (\,q_0\,,\,\,0\,,\,\,0\,,\,\,|{\bf p_2}|\,)\,,
\nonumber
\end{eqnarray}
where $E_2 = (m_1^2+m_2^2-q^2)/2 m_1$ and $q_0=(m_1^2-m_2^2+q^2)/2 m_1$.
In the $B(B_c)$-rest frame the polarization vectors of the effective 
current  read

\begin{eqnarray}
\epsilon^\mu(t)&=&
\frac{1}{\sqrt{q^2}}(\,q_0\,,\,\,0\,,\,\,0\,,\,\,|{\bf p_2}|\,)
\,,\nonumber\\
\epsilon^\mu(\pm) &=& 
\frac{1}{\sqrt{2}}(\,0\,,\,\,\mp 1\,,\,\,-i\,,\,\,0\,)\,,
\label{hel_basis}\\
\epsilon^\mu(0) &=&
\frac{1}{\sqrt{q^2}}(\,|{\bf p_2}|\,,\,\,0\,,\,\,0\,,\,\,q_0\,)\,.
\nonumber
\end{eqnarray}
Using this basis one can express the components of the hadronic
tensors through the invariant form factors defined in  Eq.~(\ref{ff}). 

\vspace*{0.5cm}
\noindent
(a) $B(B_c)\to K(D)$ transition:

\begin{equation}
\label{hel_pp}
H^{ij}(m,n)= \left(\epsilon^{\dagger \mu}(m)T^i_\mu\right)\cdot 
        \left(\epsilon^{\dagger \nu}(n)T^j_\nu\right)^\dagger\equiv
H^i(m)H^{\dagger j}(n)
\end{equation}
The helicity form factors $H^i(m)$ are given in terms of the
invariant form factors. One has

\begin{eqnarray}
H^i(t) &=& \frac{1}{\sqrt{q^2}}(Pq\, {\cal F}^i_+ + q^2\, 
{\cal F}^i_-)\,,\nonumber\\
H^i(\pm) &=& 0\,,
\label{hel_pp1}\\
H^i(0) &=& \frac{2\,m_1\,|{\bf p_2}|}{\sqrt{q^2}} \,{\cal F}^i_+ \,.
\nonumber
\end{eqnarray} 

\noindent
(b) $B_c\to D^\ast$ transition:

\begin{eqnarray} 
H^{ij}(m,n) &=&  
\epsilon^{\dagger \mu}(m) \epsilon^{ \nu}(n)H^{ij}_{\mu\nu}
=
\epsilon^{\dagger \mu}(m) \epsilon^{ \nu}(n) 
T^i_{\mu\alpha} 
\left(-g^{\alpha\beta}+\frac{p_2^\alpha p_2^\beta}{m_2^2}\right)
T^{\dagger j}_{\beta\nu}
\nonumber\\
&=&
\epsilon^{\dagger \mu}(m) \epsilon^{ \nu}(n) 
T^i_{\mu\alpha}\epsilon_2^{\dagger\alpha}(r)
\epsilon_2^{\beta}(s)\delta_{rs}
T^{\dagger j}_{\beta\nu}
\label{hel_vv}\\
&=&
\epsilon^{\dagger \mu}(m)\epsilon_2^{\dagger\alpha}(r)
T^i_{\mu\alpha} \cdot
\left(\epsilon^{\dagger \nu}(n)\epsilon_2^{\dagger\beta}(s)T^j_{\nu\beta}
\right)^\dagger\delta_{rs}
=H^i(m)H^{\dagger \,j}(n).
\nonumber
\end{eqnarray} 
From angular momentum conservation one has 
$r=m$ and $s=n$ for $m,n=\pm,0$ and $r,s=0$ for $m,n=t$.
For further evaluation one needs to specify the helicity components
$\epsilon_2(m)$ $(m=\pm,0)$ of the polarization vector of the $D^\ast$.
They read
\begin{eqnarray}
\epsilon^\mu_2(\pm) &=& 
\frac{1}{\sqrt{2}}(0\,,\,\,\pm 1\,,\,\,-i\,,\,\,0\,)\,,
\nonumber\\
&&\label{vect_pol}\\
\epsilon^\mu_2(0) &=& 
\frac{1}{m_2}(|{\bf p_2}|\,,\,\,0\,,\,\,0\,,\,\,-E_2\,)\,.
\nonumber
\end{eqnarray}
They satisfy the  orthonormality and completeness properties:
\begin{eqnarray}
&&
\epsilon_2^{\dagger\mu}(r)\epsilon_{2\mu}(s)=-\delta_{rs},
\nonumber\\
&&\\
&&
\epsilon_{2\mu}(r)\epsilon^\dagger_{2\nu}(s)\delta_{rs}=
-g_{\mu\nu}+\frac{p_{2\mu}p_{2\nu}}{m_2^2}.\nonumber
\end{eqnarray}

Finally one obtains the non-zero components of the hadron tensors 
\begin{eqnarray}
H^i(t) &=& 
\epsilon^{\dagger \mu}(t)\epsilon_2^{\dagger \alpha}(0)T^i_{\mu\alpha}
\,=\,
\frac{1}{m_1+m_2}\frac{m_1\,|{\bf p_2}|}{m_2\sqrt{q^2}}
\left(Pq\,(-A^i_0+A^i_+)+q^2 A^i_-\right),
\nonumber\\
H^i(\pm) &=& 
\epsilon^{\dagger \mu}(\pm)\epsilon_2^{\dagger \alpha}(\pm)
T^i_{\mu\alpha}
\,=\,
\frac{1}{m_1+m_2}\left(-Pq\, A^i_0\pm 2\,m_1\,|{\bf p_2}|\, V^i \right),
\label{hel_vv1}\\
H^i(0) &=&  
\epsilon^{\dagger \mu}(0)\epsilon_2^{\dagger \alpha}(0)T^i_{\mu\alpha}
\nonumber\\
&=&
\frac{1}{m_1+m_2}\frac{1}{2\,m_2\sqrt{q^2}} 
\left(-Pq\,(m_1^2+m_2^2-q^2)\, A^i_0
+4\,m_1^2\,|{\bf p_2}|^2\, A^i_+\right).\nonumber
\end{eqnarray} 

The lepton tensors $L^{(k)}(m,n)$ are evaluated 
in the $\bar ll$-CM system $\vec k_1+\vec k_2=0$.
One has (see Fig.~\ref{fig:bkangl})
\begin{eqnarray}
q^\mu   &=& (\,\sqrt{q^2}\,,\,\,0\,,\,\,0\,,\,\,0\,)\,,
\nonumber\\
k^\mu_1 &=& 
(\,E_1\,,\,\, |{\bf k_1}|\sin\theta\cos\chi\,,\,\, 
|{\bf k_1}| \sin\theta\sin\chi\,,\,\,|{\bf k_1}| \cos\theta\,)\,,
\label{kpi_basis}\\
k^\mu_2 &=& (\,E_1\,,\,\,-|{\bf k_1}|\sin\theta\cos\chi\,,\,\,
-|{\bf k_1}|\sin\theta\sin\chi\,,\,\,-|{\bf k_1}|\cos\theta\,)\,,
\nonumber
\end{eqnarray}
with $E_1=\sqrt{q^2}/2$ and $|{\bf k_1}|=\sqrt{q^2-4\mu^2}/2$.
The longitudinal and time component polarization vectors 
in the $\bar l l$ rest frame 
can be read off from Eq.~(\ref{hel_basis}) and are given by
$\epsilon^\mu(0)=(0,0,0,1)$ and $\epsilon(t)=(1,0,0,0)$ whereas the
transverse parts remain unchanged from Eq.~(\ref{hel_basis}).

The differential $(q^2,\cos\theta)$ distribution finally reads

\begin{eqnarray}
\frac{d\Gamma(H_{in}\to H_f\bar l l)}{dq^2d(\cos\theta)} &=&\,
\frac{3}{8}\,(1+\cos^2\theta)\cdot
\frac{1}{2}\left(\frac{d\Gamma_U^{11}}{d q^2}+
                   \frac{d\Gamma_U^{22}}{d q^2}\right)
\label{distr2}\\
&+&\frac{3}{4}\,\sin^2\theta\cdot
\frac{1}{2}\left(\frac{d\Gamma_L^{11}}{d q^2}+
                   \frac{d\Gamma_L^{22}}{d q^2}\right)\nonumber\\
&-&\,v \cdot\frac{3}{4}\cos\theta\cdot\frac{d\Gamma_P^{12}}{d q^2}
\nonumber\\
&+&\frac{3}{4}\,\sin^2\theta\cdot\frac{1}{2}
 \frac{d\tilde\Gamma_U^{11}}{d q^2}
-\frac{3}{8}\,(1+\cos^2\theta)\cdot\frac{d\tilde\Gamma_U^{22}}{d q^2}
\nonumber\\
&+&\frac{3}{2}\,\cos^2\theta\cdot\frac{1}{2}
 \frac{d\tilde\Gamma_L^{11}}{d q^2}
-\frac{3}{4}\,\sin^2\theta\cdot\frac{d\tilde\Gamma_L^{22}}{d q^2}
+\frac{1}{4}\,\frac{d\tilde\Gamma_S^{22}}{d q^2}\,.
\nonumber
\end{eqnarray}
Integrating over $\cos\theta$ one obtains 
\begin{eqnarray}
\frac{d\Gamma(H_{in} \to H_f \bar l l)}{dq^2} &=&\,
\frac{1}{2}
\left(
 \frac{d\Gamma_U^{11}}{d q^2}+\frac{d\Gamma_U^{22}}{d q^2}
+\frac{d\Gamma_L^{11}}{d q^2}+\frac{d\Gamma_L^{22}}{d q^2}
\right)
\label{distr1}\\
&+&\frac{1}{2}\frac{d\tilde\Gamma_U^{11}}{d q^2} 
-\frac{d\tilde\Gamma_U^{22}}{d q^2}
+\frac{1}{2}\frac{d\tilde\Gamma_L^{11}}{d q^2}
-\frac{d\tilde\Gamma_L^{22}}{d q^2}
+\frac{1}{2}\, \frac{d\tilde\Gamma_S^{22}}{d q^2}\,,
\nonumber
\end{eqnarray}
where the partial helicity rates $d\Gamma_X^{ij}/dq^2$ and 
                           $d\tilde\Gamma_X^{ij}/dq^2$
($X=U,L,P,S;\,i,j=1,2$) are defined as 

\begin{eqnarray}
\frac{d\Gamma_{X_{ij}}}{dq^2} &=& \frac{G^2_F}{(2\pi)^3} 
\left(\frac{\alpha|\lambda_t|}{2\pi}\right)^2
\frac{|{\bf p_2}|\,q^2\,v}{12\,m_1^2}\,H_X^{ij} \,,
\nonumber\\
&&\label{hel_rate1}\\
\frac{d\tilde\Gamma_{X_{ij}}}{dq^2} &=&
\frac{2\,\mu^2}{q^2}\,\frac{d\Gamma_X^{ij}}{dq^2} \,.
\nonumber
\end{eqnarray}
The relevant bilinear combinations of the helicity amplitudes
are defined in Table~\ref{tab:helicity}.

To check our calculation we give the corresponding expression  
for the differential decay rate  used in \cite{Ali:1999mm} and 
\cite{Geng:2001vy}: 
\begin{eqnarray}\label{check_1}
    \frac{d\Gamma \left( B\rightarrow K l^+l^-\right) }{ds}
        &=&\frac{G_F^2|\lambda _t|^2m_{1}^5\alpha^2}
        {3\cdot 2^9\pi ^5}v\phi_H ^{1\over 2}\left[ \left(1+\frac{2t}
        {s}\right) \phi_H \alpha _2+12t \beta _2\right]\,,
\end{eqnarray}
with  $s=q^2/m^2_1$ and $t= \mu^2/ m_{1}^2 $ and  the expressions of
$\phi_H$, $\alpha_i$, $\beta_i$,  $\delta$ (i=1,2,3) and $\delta $
are given in Appendix B. We find complete agreement with the decay 
distributions given in~\cite{Ali:1999mm} and~\cite{Geng:2001vy}.

\subsection{The four-fold angle distribution in the cascade decay \\
$B_c\to D^\ast(\to D\pi) \bar ll$.}

The lepton-hadron correlation function $L_{\mu\nu}H^{\mu\nu}$ reveals 
even more structure when one uses the cascade decay 
$B_c\to D^\ast(\to D\pi)\bar l l$ to analyze the polarization of the 
$D^\ast$. The hadron tensor now reads

\begin{equation}
H^{ij}_{\mu\nu}=T^i_{\mu\alpha}(T^j_{\nu\beta})^\dagger
\frac{3}{2\,|{\bf p_3}|}{\rm Br}(D^\ast\to D\pi)p_{3\alpha'}p_{3\beta'}
S^{\alpha\alpha'}(p_2)S^{\beta\beta'}(p_2)
\end{equation}
where 
$S^{\alpha\alpha'}(p_2)=-g^{\alpha\alpha'}+p_2^\alpha 
p_2^{\alpha'}/m_2^2$ 
is the standard spin 1 tensor,
$p_2=p_3+p_4$, $p_3^2=m_D^2$, $p_4^2=m_\pi^2$, and  $p_3$ and $p_4$
are the momenta of the $D$ and the $\pi$, respectively. 
The relative configuration of the ($D,\pi$)- and ($\bar l l$)-planes
is shown in Fig.~\ref{fig:bkangl}. 

In the rest frame of the $D^\ast$ one has 
\begin{eqnarray}\label{casc_mom}
p^\mu_2 &=& (m_{D^\ast},\vec{0}), \\
p^\mu_3 &=& (\,E_D\,,\,\,|{\bf p_3}|\,\sin\theta^\ast\,,\,\,0\,,\,\,
                    -|{\bf p_3}|\,\cos\theta^\ast\,)\,,\nonumber\\   
p^\mu_4 &=& (\,E_\pi\,,\,\,-|{\bf p_3}|\,\sin\theta^\ast\,,\,\,0\,,\,\,
                        |{\bf p_3}|\,\cos\theta^\ast\,)\, ,\nonumber\\ 
|{\bf p_3}| &=& 
\lambda^{1/2}(m_{D^\ast}^2,m_D^2,m_\pi^2)/(2\,m_{D^\ast})\, . \nonumber
\end{eqnarray}
Without loss of generality we set the azimuthal angle $\chi^\ast$
of the $(D,\pi)$-plane to zero. 
According to Eq.~(\ref{vect_pol}) the rest frame polarization vectors
of the $D^\ast$ are given by
\begin{eqnarray}\label{casc_pol}
\epsilon^\mu_2(\pm) &=& \frac{1}{\sqrt{2}}(\,0\,,\,\,\pm 1\,,\,\,
-i\,,\,\,0\,)\,,\\
\epsilon^\mu_2(0) &=& (\,0\,,\,\,0\,,\,\,0\,,\,\,-1)\,.
\nonumber
\end{eqnarray}
The spin 1 tensor $S^{\alpha\alpha'}(p_2)$ is then written as 

\begin{equation}
S^{\alpha\alpha'}(p_2)=-g^{\alpha\alpha'}
+\frac{p_2^\alpha p_2^{\alpha'}}{m_2^2}
=\sum\limits_{m=\pm,0}\epsilon_2^\alpha(m)
\epsilon_2^{\dagger\alpha'}(m) \, . 
\end{equation}
Following basically the same trick as in Eq.~(\ref{contraction})
the contraction of the lepton and hadron tensors may be
written through helicity components as
\begin{eqnarray} 
L^{(k)\mu\nu}H^{ij}_{\mu\nu}&=&
\epsilon^{\mu'}(m)\epsilon^{\dagger\nu'}(n)L^k_{\mu'\nu'}
g_{mn'}g_{nn'}\epsilon^{\dagger\mu}(m')\epsilon^\nu(n')H^{ij}_{\mu\nu}
\label{deriv}\\ 
&=&L^k(m,n)g_{mm'}g_{nn'}
\left(
\epsilon^{\dagger\mu}(m')\epsilon_2^{\dagger\alpha}(r)T^i_{\mu\alpha}
\right)
\left(
\epsilon^{\dagger\nu}(n')\epsilon_2^{\dagger\alpha}(s)T^j_{\nu\beta}
\right)^\dagger\nonumber\\
&\times& p_3\epsilon_2(r)\cdot  p_3\epsilon_2^\dagger(s)
\frac{3\,{\rm Br}(D^\ast\to D\pi)}{2\,|{\bf p_3}|}
\nonumber\\
&=&\frac{3\,{\rm Br}(D^\ast\to D\pi)}{2\,|{\bf p_3}|} 
\biggl( L^k(t,t)|H^{ij}(t)|^2\cdot (p_3\epsilon_2^\dagger(0))^2
\nonumber\\
&+&\sum\limits_{m,n=\pm,0}L^k(m,n)H^i(m)H^{\dagger j}(n)
\cdot p_3\epsilon_2(m)\cdot p_3\epsilon_2^\dagger(n)
\nonumber\\
&-&\sum\limits_{n=\pm,0}L^k(t,n)H^i(t)H^{\dagger j}(n)
\cdot p_3\epsilon_2(0)\cdot p_3\epsilon_2^\dagger(n)\nonumber\\
&-&\sum\limits_{m=\pm,0}L^k(m,t) H^i(m)H^{\dagger j}(t)
\cdot p_3\epsilon_2(m)\cdot p_3\epsilon_2^\dagger(0)
\biggr) \, . \nonumber
\end{eqnarray}

Using these results one obtains  the full four-fold angular
decay distribution 
\begin{eqnarray}
&&\frac{d\Gamma(B_c\to D^\ast(\to D\pi)\bar l l)}
     {dq^2\,d\cos\theta\,d(\chi/2\pi)\,d\cos\theta^\ast} =
{\rm Br}(D^\ast\to D\pi)
\label{distr4}\\ 
\nonumber\\ 
&&\times
\left\{
\frac{3}{8}\,(1+\cos^2\theta)\cdot\frac{3}{4}\,\sin^2\theta^\ast\cdot
\frac{1}{2}\left(\frac{d\Gamma_U^{11}}{d q^2}+
                   \frac{d\Gamma_U^{22}}{d q^2}\right)
\right.
\nonumber\\
&&
+\frac{3}{4}\,\sin^2\theta\cdot\,\frac{3}{2}\,\cos^2\theta^\ast\cdot
\frac{1}{2}\left(\frac{d\Gamma_L^{11}}{d q^2}+
                   \frac{d\Gamma_L^{22}}{d q^2}\right)\nonumber\\
&&
-\frac{3}{4}\,\sin^2\theta\cdot \cos 2\chi\cdot
 \frac{3}{4}\,\sin^2\theta^\ast\cdot
 \frac{1}{2}\left(\frac{d\Gamma_T^{11}}{d q^2}+
                  \frac{d\Gamma_T^{22}}{d q^2}\right)
\nonumber\\
&&
+\frac{9}{16}\,\sin 2\theta\cdot \cos\chi\cdot\sin 2\theta^\ast\cdot
 \frac{1}{2}\left(\frac{d\Gamma_I^{11}}{d q^2}+
                  \frac{d\Gamma_I^{22}}{d q^2}\right)
\nonumber\\
&&
+v\,
\left[-\frac{3}{4}\,\cos\theta\cdot\,\frac{3}{4}\,
\sin^2\theta^\ast\cdot\frac{d\Gamma_P^{12}}{d q^2}
\right.
\nonumber\\
&&
-\frac{9}{8}\,\sin\theta\cdot \cos\chi\cdot\sin 2\theta^\ast\cdot
 \frac{1}{2}\left(\frac{d\Gamma_A^{12}}{d q^2}+
                    \frac{d\Gamma_A^{21}}{d q^2}\right)
\nonumber\\
&&
\left.
+\frac{9}{16}\,\sin \theta\cdot\sin\chi\cdot\sin 2\theta^\ast\cdot
\left(\frac{d\Gamma_{II}^{12}}{d q^2}
+\frac{d\Gamma_{II}^{21}}{d q^2}\right)\right]\nonumber\\
&&
-\frac{9}{32}\,\sin 2\theta\cdot\sin\chi\cdot\sin 2\theta^\ast\cdot 
\left(\frac{d\Gamma_{IA}^{11}}{d q^2}+\frac{d\Gamma_{IA}^{22}}{d q^2}
\right)\nonumber\\
&&
+\frac{9}{32}\,\sin^2\theta\cdot\sin 2\chi\cdot\sin^2\theta^\ast\cdot 
\left(\frac{d\Gamma_{IT}^{11}}{d q^2}+\frac{d\Gamma_{IT}^{22}}{d q^2}
\right)\nonumber\\
&&
+\frac{3}{4}\,\sin^2\theta\cdot\frac{3}{4}\,\sin^2\theta^\ast\cdot
 \frac{1}{2}\cdot\frac{d\tilde\Gamma_U^{11}}{d q^2}
 - \frac{3}{8}\,(1+\cos^2\theta)\cdot\frac{3}{4}\,\sin^2\theta^\ast\cdot
   \frac{d\tilde\Gamma_U^{22}}{d q^2}
\nonumber\\
&&
+\frac{3}{2}\,\cos^2\theta\cdot\frac{3}{2}\,\cos^2\theta^\ast\cdot
 \frac{1}{2}\cdot\frac{d\tilde\Gamma_L^{11}}{d q^2}
-\frac{3}{4}\,\sin^2\theta\cdot\frac{3}{2}\,\cos^2\theta^\ast\cdot
    \frac{d\tilde\Gamma_L^{22}}{d q^2}
\nonumber\\
&&
+\frac{3}{4}\,\sin^2\theta\cdot \cos 2\chi\cdot
 \frac{3}{4}\,\sin^2\theta^\ast\cdot
 \left(\frac{d\tilde\Gamma_T^{11}}{d q^2}+
       \frac{d\tilde\Gamma_T^{22}}{d q^2}\right)
\nonumber\\
&&
-\frac{9}{8}\,\sin 2\theta\cdot \cos\chi\cdot \sin 2\theta^\ast\cdot
  \frac{1}{2}\left(\frac{d\tilde\Gamma_I^{11}}{d q^2}+
                     \frac{d\tilde\Gamma_I^{22}}{d q^2}\right)
+\frac{3}{2}\,\cos^2\theta^\ast\cdot \frac{1}{4}
 \frac{d\tilde\Gamma_S^{22}}{d q^2}
\nonumber\\
&&
+\frac{9}{16}\,\sin 2\theta\cdot\sin\chi\cdot\sin 2\theta^\ast\cdot 
\left(\frac{d\tilde\Gamma_{IA}^{11}}{d q^2}
+\frac{d\tilde\Gamma_{IA}^{22}}{d q^2}\right)
\nonumber\\
&&
\left.
-\frac{9}{16}\,\sin^2\theta\cdot\sin 2\chi\cdot\sin^2\theta^\ast\cdot 
\left(\frac{d\tilde\Gamma_{IT}^{11}}{d q^2}
+\frac{d\tilde\Gamma_{IT}^{22}}{d q^2}\right)
\nonumber
\right\}
\end{eqnarray} 
Integrating Eq.~(\ref{distr4}) over $\cos\theta^\ast$ and 
$\chi$ one recovers the two-fold ($q^2,\cos\theta$) distribution of
Eq.~(\ref{distr2}). 
Note that a similar four-fold distribution has also been obtained in 
Refs.(\cite{Kruger:1999xa},\cite{Kim:2000dq}-\cite{Melikhov:1998cd}) 
using, however, the zero lepton mass approximation. 
If there are sufficient data one can attempt to fit them to the
full four-fold decay distribution and thereby extract the values
of the coefficient functions $d\Gamma_X/dq^2$ and, in the case $l=\tau$
the coefficient functions $d\tilde\Gamma_X/dq^2$.
Instead of considering the full four-fold decay distribution
one can analyze single angle distributions by integrating out
two of the remaining angles as e.g. discussed in Sec.3.2
Observables related to single angle distributions will be discussed
in the next subsection.

\subsection{Physical observables}

The four-fold distribution Eq.~(\ref{distr4}) allows one to define 
a number of physical observables which can be measured
experimentally.
An asymmetry parameter $\alpha_{\theta^\ast}$ is defined from the
angular distribution 
$W(\cos^2\theta^\ast)=1+\alpha_{\theta^\ast}\cos^2\theta^\ast$.
Integrating Eq.~(\ref{distr4}) over $\cos\theta$ and $\chi$ one finds

\begin{equation}
\label{as_star}
\alpha_{\theta^\ast}=\frac{
-(U^{11}+U^{22})+2(L^{11}+L^{22})-\tilde U^{11}+2\tilde U^{22}
+2(\tilde L^{11}-2\tilde L^{22})+2\tilde S^{22}}
{U^{11}+U^{22}+\tilde U^{11}-2\tilde U^{22}}
\end{equation}
By integrating over $\cos\theta^\ast$ and $\chi$ 
one can define two asymmetry  parameters  $\alpha^\prime_{\theta}$ and 
$\alpha_{\theta}$ according to the angular distribution 
$W(\cos^2\theta)=1+\alpha^\prime_{\theta}\cos\theta
+\alpha_{\theta}\cos^2\theta$. One has

\begin{eqnarray}
\alpha^\prime_{\theta} &=& \frac{
-4\,v\,P^{12}}
{U^{11}+U^{22}+2(L^{11}+L^{22})+2(\tilde U^{11}-\tilde U^{22}
-2\tilde L^{22}+(2/3)\tilde S^{22})}\,,
\label{as_theta1}\\
\alpha_{\theta} &=& \frac
{U^{11}+U^{22}-2(L^{11}+L^{22})-2(\tilde U^{11}+\tilde U^{22})
+4(\tilde L^{11}+\tilde L^{22})}
{U^{11}+U^{22}+2(L^{11}+L^{22})+2(\tilde U^{11}-\tilde U^{22}
-2\tilde L^{22}+(2/3)\tilde S^{22})}\,.
\label{as_theta2}
\end{eqnarray} 

An azimuthal asymmetry parameter $\beta$ can be  defined from 
the $\chi$-distribution \\
$W(\chi)=1+\beta \cos 2\chi$.
The azimuthal $\chi$-distribution is obtained by integrating 
over $\cos\theta$ and $\cos\theta^\ast$. One has

\begin{equation}
\label{as_beta}
\beta=\frac
{-(T^{11}+T^{22})+2(\tilde T^{11}+\tilde T^{22})}
{U^{11}+U^{22}+L^{11}+L^{22}+\tilde U^{11}-2\tilde U^{22}
+\tilde L^{11}-2\tilde L^{22}+\tilde S^{22}}\,.
\end{equation}

A second strategy is to define suitable asymmetry ratios that
project out the partial rates from Eq.~(\ref{distr4}).
Let us consider the following four asymmetry ratios
which project out the contributions of the parity conserving partial 
rates $\Gamma_T$ and $\Gamma_I$ and the parity violating partial
decay rates $\Gamma_P$ and $\Gamma_A$. One has
\begin{equation}
\label{as_a}
\Gamma_T: \hspace{1cm} 
A_T=\frac
{
d\Gamma(\chi)-d\Gamma(\chi+\pi/2)+d\Gamma(\chi+\pi)-d\Gamma(\chi+3\pi/2) 
}
{
d\Gamma(\chi)+d\Gamma(\chi+\pi/2)+d\Gamma(\chi+\pi)+d\Gamma(\chi+3\pi/2) 
}
\end{equation}
where $-\pi/4\le\chi\le\pi/4$.

\begin{equation}
\Gamma_A: \hspace{1cm} 
A_I = N_I/D_I\,,
\label{as_i}\\
\end{equation}

\begin{eqnarray*}
N_I &=& d\Gamma(\theta,\theta^\ast,\chi)
       -d\Gamma(\theta,\theta^\ast,\chi+\pi)\\
&-&d\Gamma(\theta,\pi-\theta^\ast,\chi)
  +d\Gamma(\theta,\pi-\theta^\ast,\chi+\pi)
\nonumber\\
&-&d\Gamma(\pi-\theta,\theta^\ast,\chi)
  +d\Gamma(\pi-\theta,\theta^\ast,\chi+\pi)
\nonumber\\
&+&d\Gamma(\pi-\theta,\pi-\theta^\ast,\chi)
  -d\Gamma(\pi-\theta,\pi-\theta^\ast,\chi+\pi)\,,
\nonumber\\
&&\nonumber\\
&&
0\le \theta^\ast\le\frac{\pi}{2}\,, \hspace{1cm}
\frac{\pi}{2}\le\theta\le\pi\,,   \hspace{1cm}
-\frac{\pi}{2}\le\chi\le\frac{\pi}{2}\,.
\nonumber
\end{eqnarray*}
The denominator $D_I$ is given by the same expression
with plus signs everywhere.

\begin{equation}
\label{as_fb}
\Gamma_P: \hspace{1cm}
A_{FB} = \frac{d\Gamma(\theta)-d\Gamma(\pi-\theta)}
                {d\Gamma(\theta)+d\Gamma(\pi-\theta)}\,,
\hspace{1cm}
\frac{\pi}{2}\le \theta\le\pi\,, 
\end{equation}

\begin{eqnarray}
\Gamma_A:\, A_{A} &=&
\frac{d\Gamma(\theta^\ast,\chi)-d\Gamma(\theta^\ast,\chi+\pi)
     -d\Gamma(\pi-\theta^\ast,\chi)+d\Gamma(\pi-\theta^\ast,\chi+\pi)}
     {d\Gamma(\theta^\ast,\chi)+d\Gamma(\theta^\ast,\chi+\pi)
     +d\Gamma(\pi-\theta^\ast,\chi)+d\Gamma(\pi-\theta^\ast,\chi+\pi)}\,,
\nonumber\\
&&\nonumber\\
&&
0\le \theta^\ast\le\frac{\pi}{2}\,, \hspace{1cm}
-\frac{\pi}{2}\le\chi\le\frac{\pi}{2}\,.
\label{as_aa}
\end{eqnarray}
We have used a notation where the angles that do not appear 
in the arguments of the differential rates $d\Gamma$
have been integrated out over their physical ranges
($0\le\theta\,(\theta^\ast)\le\pi$, $0\le\chi\le 2\pi$).
Integrating over the remaining variables (numerator and
denominator separately!) we finally obtain

\begin{eqnarray}
<A_T> &=& \frac{2}{\pi}\,
\frac
{
-(1/2)(T^{11}+T^{22})+\tilde T^{11}+\tilde T^{22}
}
{\Gamma}\,,
\label{as_tt}\\
&&\nonumber\\
<A_I> &=& \frac{2}{\pi}\,
\frac
{
(1/2)(I^{11}+I^{22})- \tilde I^{11}-\tilde I^{22}
}
{\Gamma}\,,
\label{as_ii}\\
&&\nonumber\\
<A_{FB}> &=& \frac{3\,v}{4}\,
\frac
{
P^{12}
}
{\Gamma}\,,
\label{as_fbfb}\\
&&\nonumber\\
<A_{A}> &=& \frac{3\,v}{4}\,
\frac
{
A^{12}+A^{21}
}
{\Gamma}\,.
\label{as_aaa}
\end{eqnarray}

\subsection{Longitudinal polarization of the lepton 
in $B\to K \bar ll$-decay.}

Our aim is to study lepton polarization effects in 
$B\to K \bar ll$-decay. The longitudinal polarization of the final 
lepton $l^-(k_2)$ is defined by
\begin{equation}
\label{long_pol}
P^{(l)}=\frac{ d\Gamma(s_2)/dq^2 - d\Gamma(-s_2)/dq^2}
                  { d\Gamma(s_2)/dq^2 + d\Gamma(-s_2)/dq^2}
\end{equation}
where the longitudinal component of the rest frame polarization vector 
of the $l^-(k_2)$ is given by 

\begin{equation}
s_2 = \left(0,\frac{{\bf k_2}}{|{\bf k_2}|}\right)\,.
\end{equation}
Contrary to previous studies of the longitudinal polarization 
of the lepton, where the longitudinal polarization was studied
in the B rest frame, we will calculate  this quantity in 
the $\bar ll$-CM frame. The longitudinal  polarization vector  $s_2$
is boosted to the moving  frame by a Lorentz transformation.
One obtains
\begin{equation}
s_{2,CM} = \left(\frac{|{\bf k_2}|}{\mu},\,
\frac{E_2}{\mu}\frac{{\bf k_2}}{|{\bf k_2}|}\right)\,.
\end{equation}
The quantity $ d\Gamma(s_2)/dq^2$ in Eq.~(\ref{long_pol}) 
may be obtained from Eq.~(\ref{distr}) by the replacement 
$$
(\not\! k_2+\mu) \to \frac{1}{2}(1+\gamma_5 \not\! s_2)\,
(\not\! k_2+\mu)\,.
$$ 
Integrating  the numerator and denominator in Eq.~(\ref{long_pol})
over $\cos\theta$, one finds
\begin{equation}
\label{long_pol1}
P^{(l)} = \frac{v\,\left(dU^{12}+dL^{12}\right)}{d\Gamma}
\end{equation}
where we have adopted a short hand notation $U:=\Gamma_U$,  
$\tilde U:=\tilde\Gamma_U$, etc..
Because of CP-invariance and because the longitudinal polarization is 
a pseudoscalar quantity, the longitudinal polarization of the 
antilepton is equal and opposite to  the longitudinal polarization 
of the lepton, i.e. $P^{(l)}(l^+)=-P^{(l)}(l^-)$.

\section{Model form factors}
\label{sec:mod}

We will employ the relativistic constituent quark model 
\cite{Ivanov:1996pz,Ivanov:1999ic} to calculate the form factors 
relevant to the decays $B\to K \bar ll$ and $B_c\to D(D^\ast) \bar ll$. 
This model is based on an effective interaction Lagrangian which 
describes the coupling between hadrons and their constituent quarks.

For example, the coupling of the meson $H$ to its constituent quarks  
$q_1$ and $\bar q_2 $ is given by the Lagrangian 
\begin{equation}
\label{lag}
{\cal L}_{{\rm int}} (x)=g_H H(x) \int\!\! dx_1 \!\!\int\!\! dx_2
F_H (x,x_1,x_2) \bar q(x_1) \Gamma_H \lambda_H q(x_2)\,.
\end{equation}
Here, $\lambda_H$ and $\Gamma_H$ are  Gell-Mann and Dirac
matrices  which entail the flavor and spin quantum numbers
of the meson field $H(x)$. The function $F_H$ is related to the scalar
part of the Bethe-Salpeter amplitude and characterizes the finite size
of the meson. The function $\Phi_H$ must be invariant under 
the translation  $F_H(x+a,x_1+a,x_2+a)=F_H(x,x_1,x_2) $.

In our previous papers we have used the so-called impulse approximation 
for the evaluation of the Feynman diagrams. In the impulse approximation
one omits a possible dependence of the vertex functions on external 
momenta. The impulse approximation therefore entails a certain dependence
on how loop momenta are routed through the diagram at hand.
This problem no longer exists in the present full treatment where
the impulse approximation is no longer used.
In the present calculation  we will use a particular form of 
the vertex function given by
\begin{equation}
\label{vertex}
F_H(x,x_1,x_2)= 
\delta\biggl(x-\frac{m_1x_1+m_2x_2}{m_1+m_2}\biggr)\Phi_H((x_1-x_2)^2).
\end{equation}
where $m_1$ and $m_2$ are the constituent quark masses. 
The vertex function $F_H$ evidently satisfies
the above translational invariance condition. As mentioned before
we no longer use the impulse approximation in the present calculation.

\begin{figure}[t]
 \vspace*{18cm}
\includegraphics{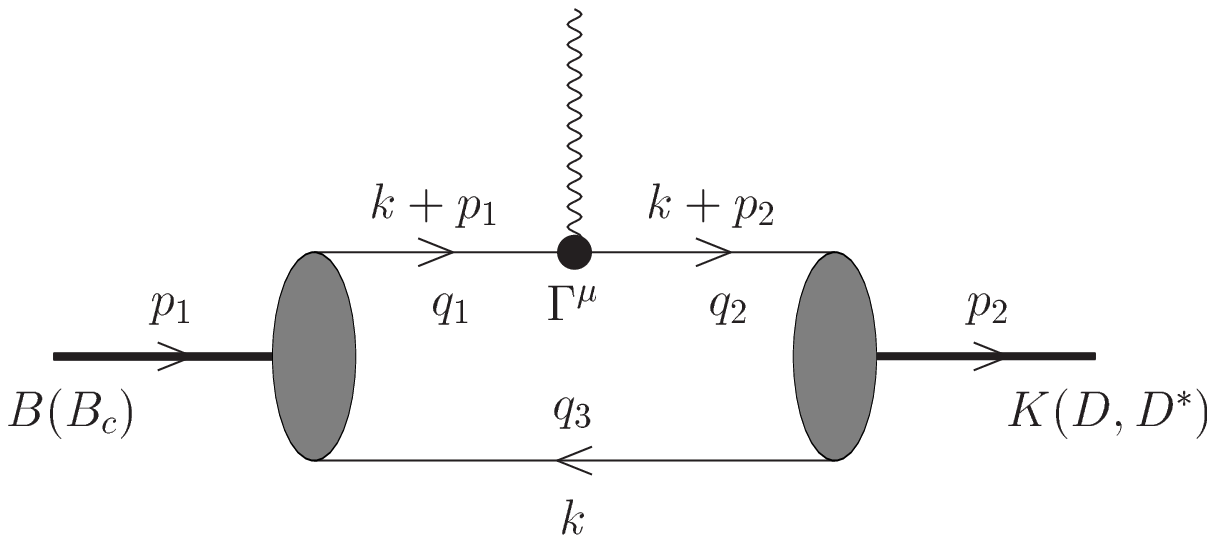}
\vspace*{-12.5cm}
\caption{Diagram describing the form factors
of the decay  $B(B_c)\to K (D, D^\ast) \bar l l$}
\label{fig:bkformf}
\end{figure}

The coupling constants $g_H$ in Eq.~(\ref{lag}) are determined  
by the so called {\it compositeness condition} proposed in \cite{SWH} 
and extensively used in \cite{Efimov:zg}. The compositeness condition 
means that the renormalization constant of the meson field is set equal 
to zero
\begin{equation}
\label{z=0}
Z_H=1-\frac{3g^2_H}{4\pi^2}\tilde\Pi^\prime_H(m^2_H)=0
\end{equation}
where $\tilde\Pi^\prime_H$ is the derivative of the meson mass operator.
For the pseudoscalar and vector mesons treated in this paper
one has
\begin{eqnarray*}
\tilde\Pi'_P(p^2)&=& \frac{1}{2p^2}\,p^\alpha\frac{d}{dp^\alpha}\,
\int\!\! \frac{d^4k}{4\pi^2i} \tilde\Phi^2_P(-k^2)
\\ 
&\times&{\rm tr} \biggl[\gamma^5 S_1(\not\! k+w_{21} \not\!p) \gamma^5 
                         S_2(\not\! k-w_{12} \not\!p) \biggr]\\
&&\\
\tilde\Pi'_V(p^2)&=&
\frac{1}{3}\biggl[g^{\mu\nu}-\frac{p^\mu p^\nu}{p^2}\biggr] 
\frac{1}{2p^2}\,p^\alpha\frac{d}{dp^\alpha}\,
\int\!\! \frac{d^4k}{4\pi^2i} \tilde\Phi^2_V(-k^2)
\\
&\times&
{\rm tr} \biggl[\gamma^\nu S_1(\not\! k+w_{21} \not\!p) \gamma^\mu 
                           S_2(\not\! k-w_{12}\not\! p)\biggr]
\end{eqnarray*}
where $w_{ij}=m_j/(m_i+m_j)$, $\tilde\Phi_H(-k^2)$ is the 
Fourier-transform of the correlation function $\Phi_H((x_1-x_2)^2)$ 
and $S_i(\not\! k)$ is the quark propagator. 
The leptonic decay constant $f_P$ is calculated from
\begin{equation}
\label{leptonic}
\frac{3g_P}{4\pi^2} \,\int\!\! \frac{d^4k}{4\pi^2i} 
\tilde\Phi_P(-k^2) 
{\rm tr} \biggl[O^\mu S_1(\not\! k+w_{21} \not\!p) \gamma^5 
                        S_2(\not\! k-w_{12} \not\!p) \biggr]
=f_P\,p^\mu.
\end{equation}
The transition form factors  $P(p_1)\to P(p_2),V(p_2)$ 
 can be calculated from the Feynman integral
corresponding to the diagram of Fig.~\ref{fig:bkformf}:
\begin{eqnarray}
\Lambda^{\Gamma^\mu}(p_1,p_2)&=&
\frac{3g_P g_{P'(V)}}{4\pi^2} \,\int\!\! \frac{d^4k}{4\pi^2i}
\tilde\Phi_P(-(k+w_{13}\,p_1)^2)\,
\tilde\Phi_{P'(V)}(-(k+w_{23}\,p_2)^2)\nonumber\\
&\times&
{\rm tr} \biggl[S_2(\not\! k+\not\! p_2) \Gamma^\mu 
 S_1(\not\! k+\not\! p_1)\gamma^5 S_3(\not\! k)\Gamma_{\rm out}\biggr]
\label{triangle}
\end{eqnarray}
where 
$\Gamma^\mu=\gamma^\mu,\,\gamma^\mu\gamma^5,\,$
$i\sigma^{\mu\nu}q_\nu,\,$ or
$i\sigma^{\mu\nu}q_\nu\gamma^5 $ and 
$\Gamma_{\,P',V}=\gamma^5,\,\gamma_\nu\epsilon_2^\nu. $

We use the local quark propagators
\begin{equation}
S_i(\not\! k)=\frac{1}{m_i-\not\! k} \,,
\end{equation}
where $m_i$ is  the constituent quark mass. 
We do not introduce a new notation for constituent quark masses
in order to distinguish  them from the current quark masses used 
in the effective Hamiltonian and Wilson coefficients as described 
in Sec. II because it should always be clear from the context which
set of masses is being referred to.
As discussed in \cite{Ivanov:1996pz,Ivanov:1999ic}, we assume that 
\begin{equation}
\label{conf}
m_H<m_{1}+m_{2}
\end{equation}
in order to avoid the appearance of imaginary parts in  the physical
amplitudes. 

The fit values for the constituent quark masses are taken from our
papers \cite{Ivanov:1996pz,Ivanov:1999ic} and are given in 
Eq.~(\ref{fitmas}). 
\begin{equation}
\begin{array}{ccccc}
m_u & m_s & m_c & m_b &   \\
\hline
$\ \ 0.235\ \ $ & $\ \ 0.333\ \ $ & $\ \ 1.67\ \ $ & $\ \ 5.06\ \ $ & 
$\ \ {\rm GeV} $\\
\end{array}
\label{fitmas}
\end{equation}
It is readily seen that the constraint Eq.~(\ref{conf}) holds true for 
the low-lying flavored pseudoscalar mesons but is no longer true
for the vector mesons.  
In the case of the heavy mesons $D^\ast$ and $B^\ast$ we
will employ identical  masses for the vector mesons
and the pseudoscalar mesons for the  calculation of
matrix elements in Eqs.~(\ref{z=0}),(\ref{leptonic}) and 
(\ref{triangle}).
It is a quite reliable approximation  because of
$(m_{D^\ast}-m_D)/m_D\sim 7\%$ and 
$(m_{B^\ast}-m_B)/m_B\sim 1\%$.  

We employ a Gaussian for the vertex function 
$\tilde\Phi_H(k^2_E/\Lambda^2_H) = \exp(-k^2_E/\Lambda^2_H)$ 
where $k_E$ is the Euclidean momentum and determine the size 
parameters $\Lambda_H$ by a fit to the experimental data, 
when available, or to lattice simulations for the leptonic decay 
constants. The quality of the fit can be seen from 
Table~\ref{tab:fit}. The branching ratios of the semileptonic 
decays are shown in Table~\ref{tab:sem}. 
The numerical values for $\Lambda_H$ are  
$\Lambda_\pi=1$ GeV, $\Lambda_K=1.6$ GeV, 
$\Lambda_D=2$ GeV and $\Lambda_B=2.25$ GeV for 
all $K$, $D$ and $B$ partners, respectively. 

\begin{figure} 
 \vspace*{5cm}
\includegraphics{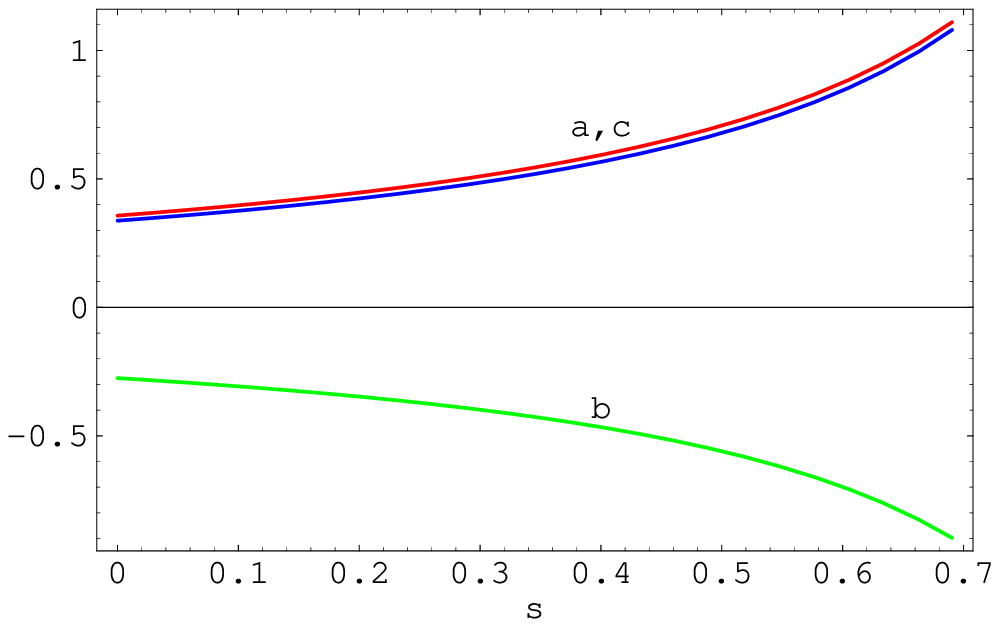}
 \vspace*{2.5cm}
\caption{
Form factors for the $B\to K$ transition: 
$(a)\,  F_+$, $(b)\,  F_-$, $(c)\,  F_T$.
}
\label{fig:bkp}

\vspace*{25cm}
\includegraphics{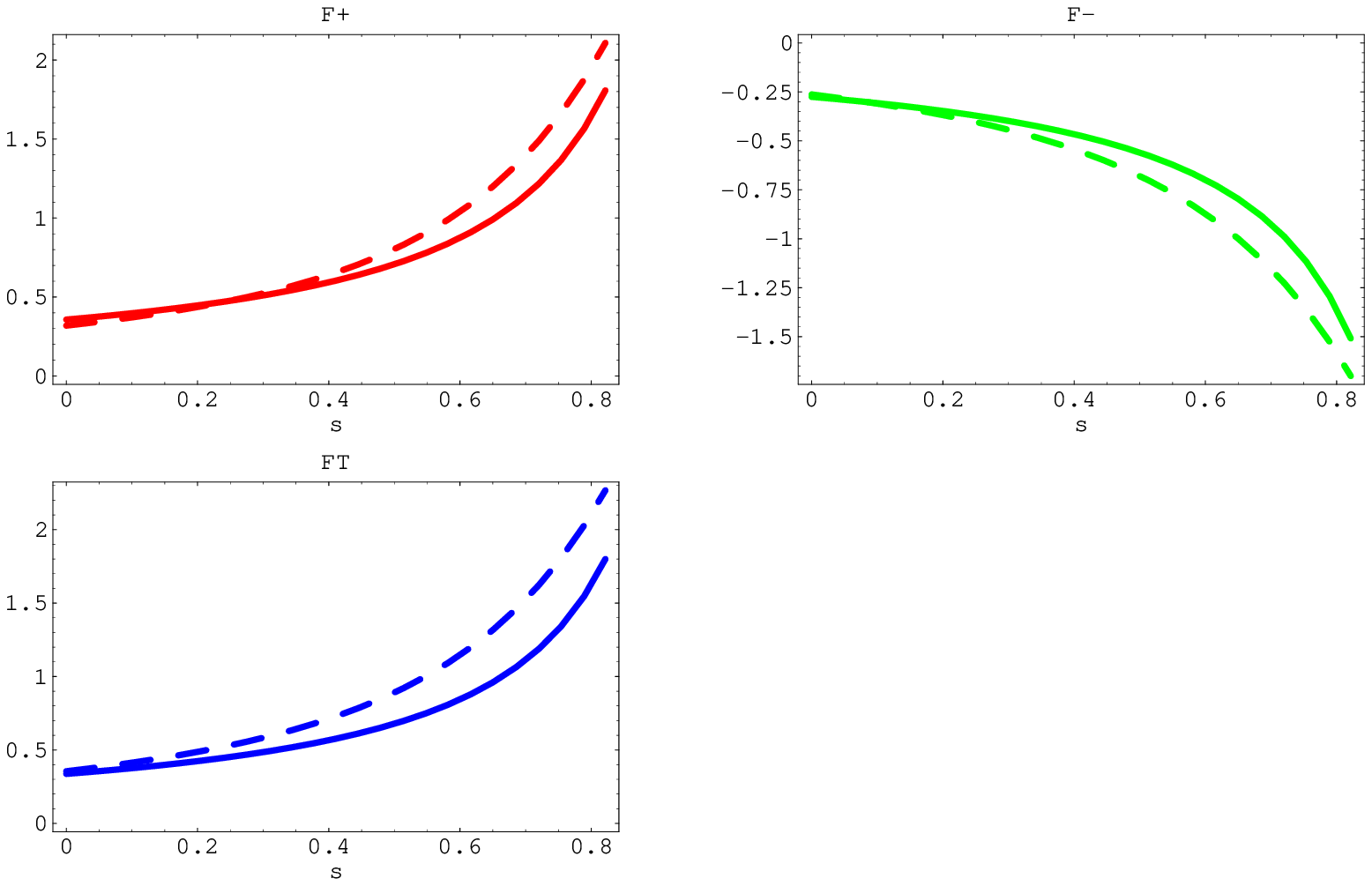}
\vspace*{-15cm}
\caption{
Comparison of our  $B\to K$ form factors (solid line) with
those used in~\cite{Ali:1999mm} (dashed line). }
\label{fig:bkp_comp}
\end{figure}

We are now in a position to present our results for the
$B\to K$ form factors. We have used the technique outlined 
in our previous papers \cite{Ivanov:1996pz,Ivanov:1999ic}
for the numerical  evaluation  of the Feynman integrals 
in Eq.~(\ref{triangle}).
The results of our numerical calculations are well represented
by the parametrization
\begin{equation}
F(s)=\frac{F(0)}{1-a s+b s^2}\,.
\end{equation} 
Using such a  parametrization facilitates further integrations.
The values of $F(0)$, $a$ and $b$ are listed  in 
Tables~\ref{tab:ffbk}-\ref{tab:ffbcd}. We plot our form factors 
in Fig.~\ref{fig:bkp} and compare them with those used in 
paper \cite{Ali:1999mm} in Fig.~\ref{fig:bkp_comp}. 
The functional behavior of the curves is similar to each other. 

At the end of this section we would like to discuss 
the impulse approximation used in our previous papers
\cite{Ivanov:1996pz,Ivanov:1999ic}. It was simply assumed
that the vertex functions depend only on the loop momentum
flowing through the vertex. The explicit translational
invariant vertex function in Eq.~(\ref{vertex}) allows one
to check the reliability of this approximation. 
We found that the results obtained with and without the impulse
approximation are rather close to each other except for the
heavy-to-light form factors. We consider 
the $B\to\pi$-transition as an example to illustrate this point.
The calculated values of the $F_+^{B\pi}(q^2)$ form factor
at $q^2=0$ are  
\[
F_+^{B\pi}(0)=\left\{  
\begin{array}{ll} 0.27 & \mbox{exact} \\
& \\
 0.48 & \mbox{impulse approximation}
\end{array}
\right.
\]
One can see that the value of the form factor at  $q^2=0$ calculated 
without the impulse approximation is considerably smaller than when 
calculated with the impulse approximation.
Its value is  close to the value of QCD SR estimates,
see, for example, \cite{Bagan:1997bp}:  

\[
F_+^{B\pi}(0)=\left\{  
\begin{array}{ll} 0.25  & \mbox{asymptotic distribution} \\
& \\
0.30  & \mbox{QCD SR distribution }
\end{array}
\right.
\]

\section{Numerical results}
\label{sec:res}

\begin{figure}
\vspace*{19cm}  
\includegraphics{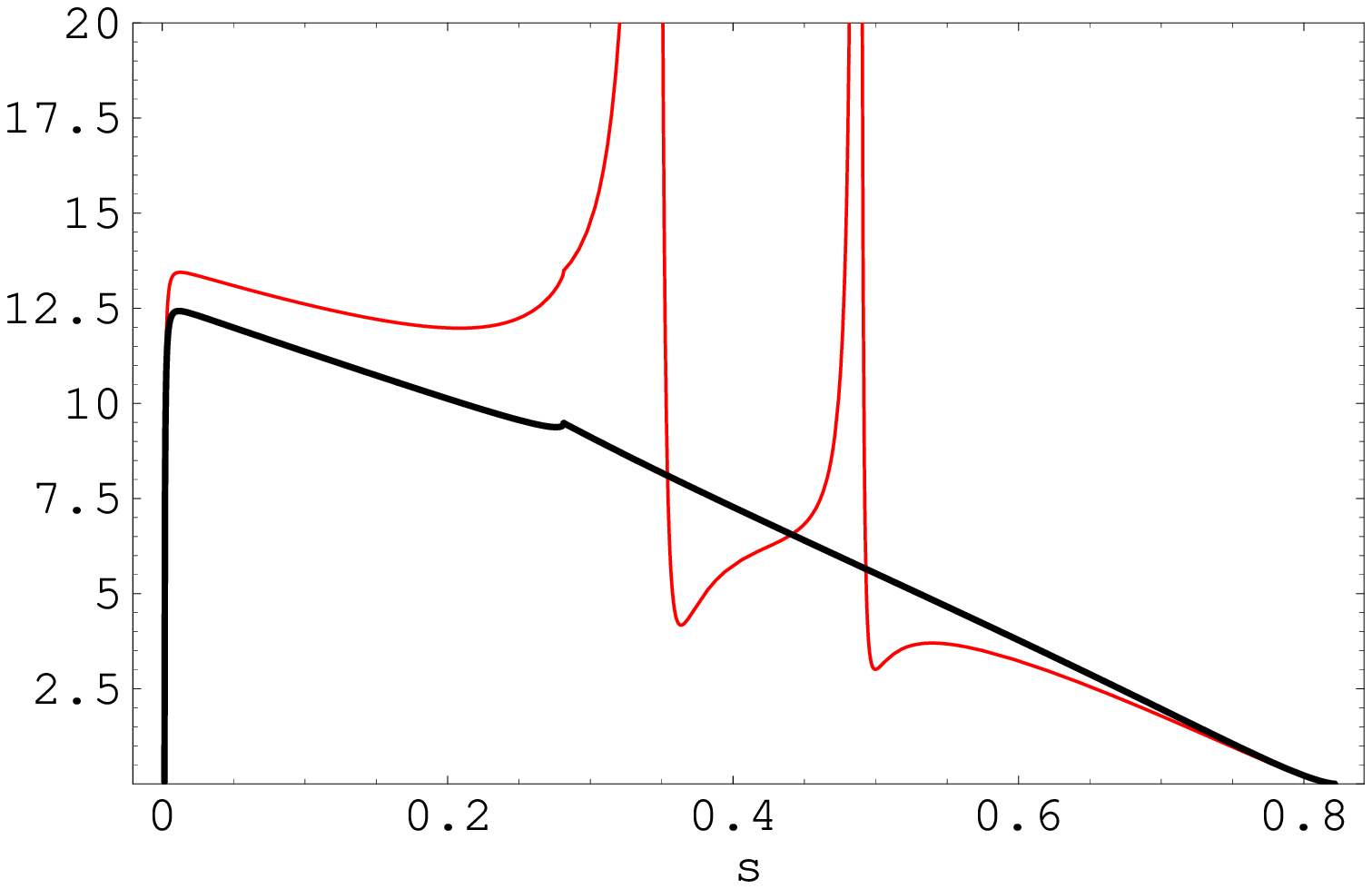}
\vspace*{-11cm}
\caption{
Normalized differential distributions 
$10^7\,\Gamma^{-1}_{\rm tot} d{\Gamma}/ds$ 
for $B\to K\, \mu^+\mu^-$. 
The curves with resonant shapes represent long-distance
contributions.}
\label{fig:bk_ll}

\vspace*{21cm} 
\includegraphics{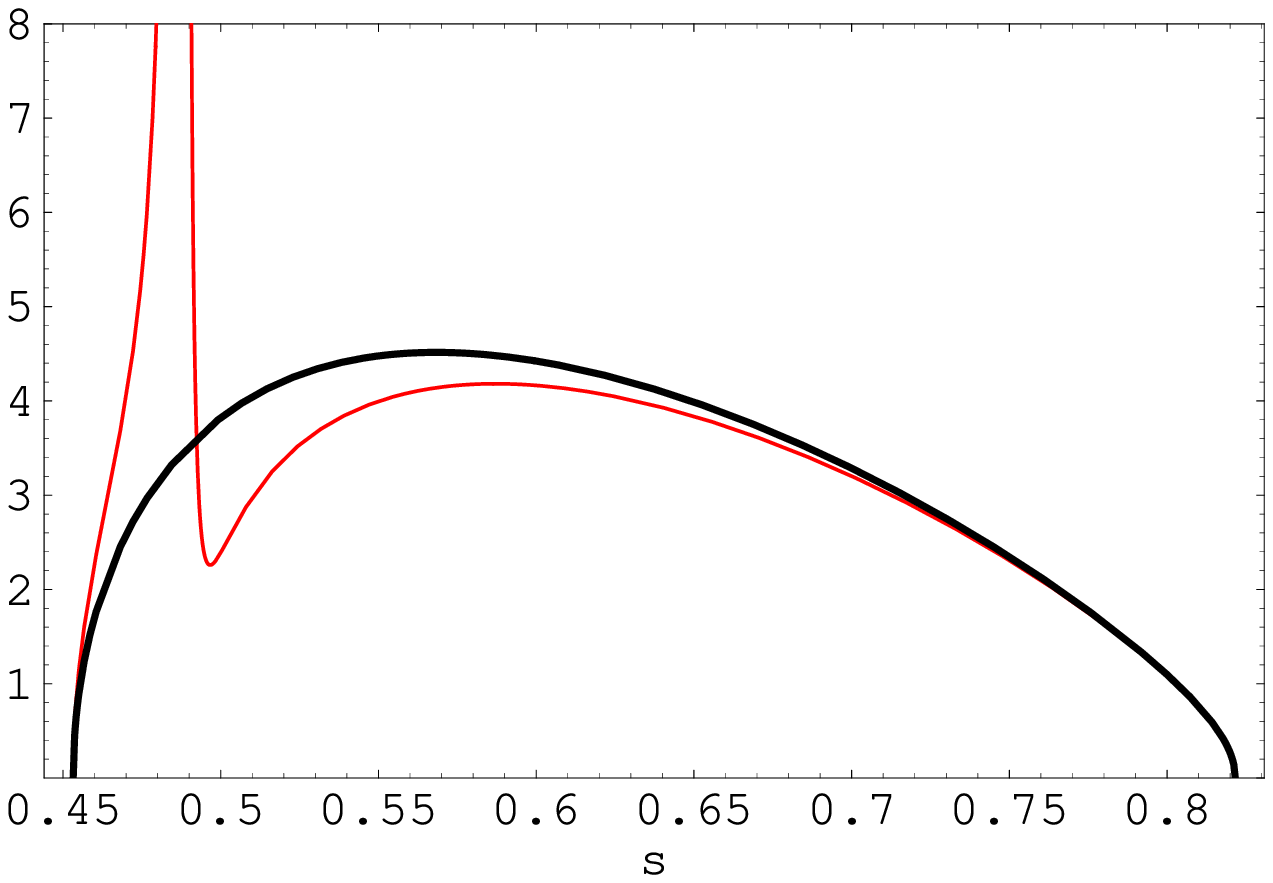}
\vspace*{-12cm} 
\caption{
Normalized differential distributions
$10^7\,\Gamma^{-1}_{\rm tot} d{\Gamma}/ds$ 
for $B\to K\, \tau^+\tau^-$.}
\label{fig:bk_tau}
\end{figure}

In this section we collect and discuss our numerical results. 
We plot the normalized differential distributions 
$\Gamma^{-1}_{\rm tot}\,d\Gamma/ds$ with 
$\Gamma_{\rm tot}=1/\tau_B$ 
($\tau_B=(\tau_{B^0}+\tau_{B^+})/2=1.60\,{\rm ps}$)
and $s=q^2/m^2_B$ in Figs.~\ref{fig:bk_ll}-\ref{fig:bk_tau} 
for the decay $B\to K \bar ll$. 

\begin{figure}
\vspace*{19.5cm}
\includegraphics{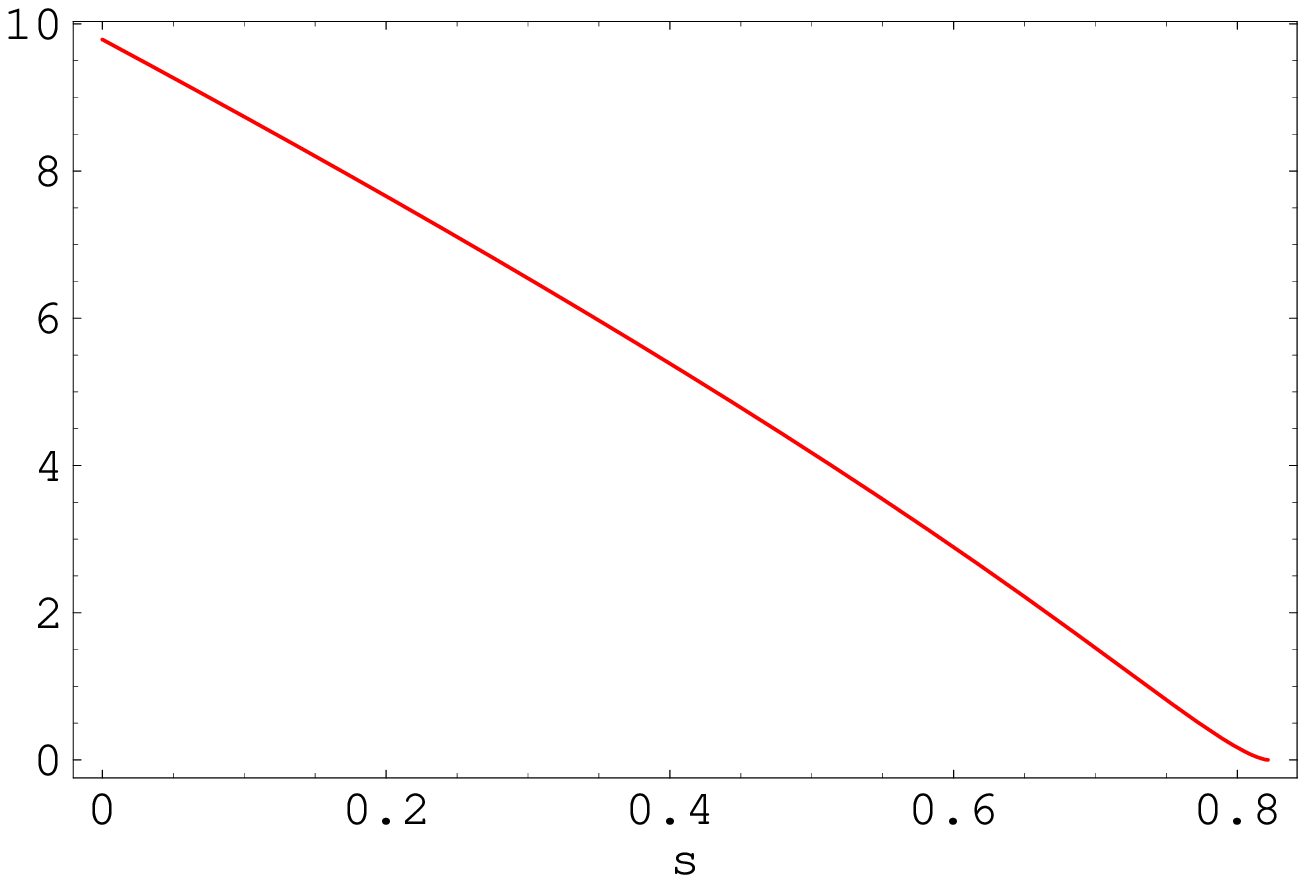}
\vspace*{-12cm}
\caption{
Normalized differential distributions 
$10^8\,\Gamma^{-1}_{\rm tot} d{\Gamma}/ds$ 
for $B\to K\,\bar\nu\nu$. 
}
\label{fig:bk_nn}


\vspace*{7.5cm}
\includegraphics{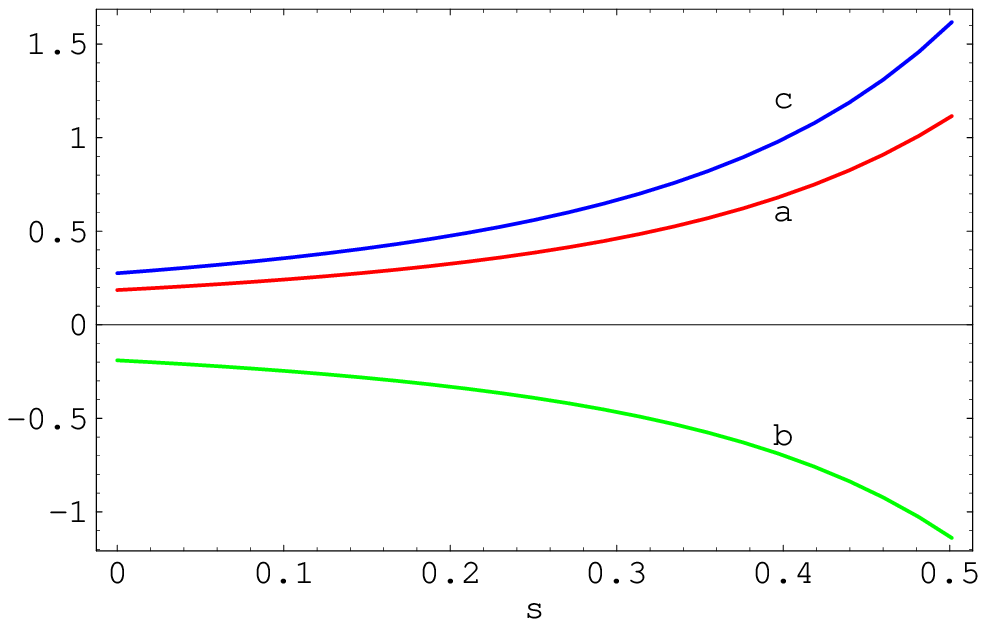}
\vspace*{1.5cm}
\caption{
Form factors for the $B_c\to D_d$ transition: 
$(a)\, F_+$, $(b)\, F_-$, $(c)\, F_T$.
}
\label{fig:bc_d}
\end{figure}

We have also included the $B\to K \bar\nu\nu$ modes. 
Their differential rates are calculated according to

\begin{eqnarray}\label{check_2} 
\frac{d\Gamma \left( B^+\rightarrow K  \bar\nu\nu\right) }{ds}
       &=&\frac{G_F^2 \, m_{B}^5 \, |\lambda_t|^2
       \alpha^2 \, |D_\nu\left( x_{t}\right)|^2}{2^8 \, \pi ^5 \, 
       \sin^4\theta_W} \, \left|F_+\right|^2 \, \phi_H^{3\over 2}\,.  
\label{neutrino} 
\end{eqnarray}
The functions $D_\nu(x_t)$ and $\phi_H$ are given in Appendix B. 
The behavior of the normalized differential distributions 
is shown in Fig.~\ref{fig:bk_nn}. 

We list our numerical results for the branching ratios in 
Table~\ref{tab:branching}. 
When comparing the values of the branching ratios with 
those obtained in \cite{Ali:1999mm}
one finds that they almost agree with each other. 

\begin{figure}
\vspace*{19.5cm} 
\includegraphics{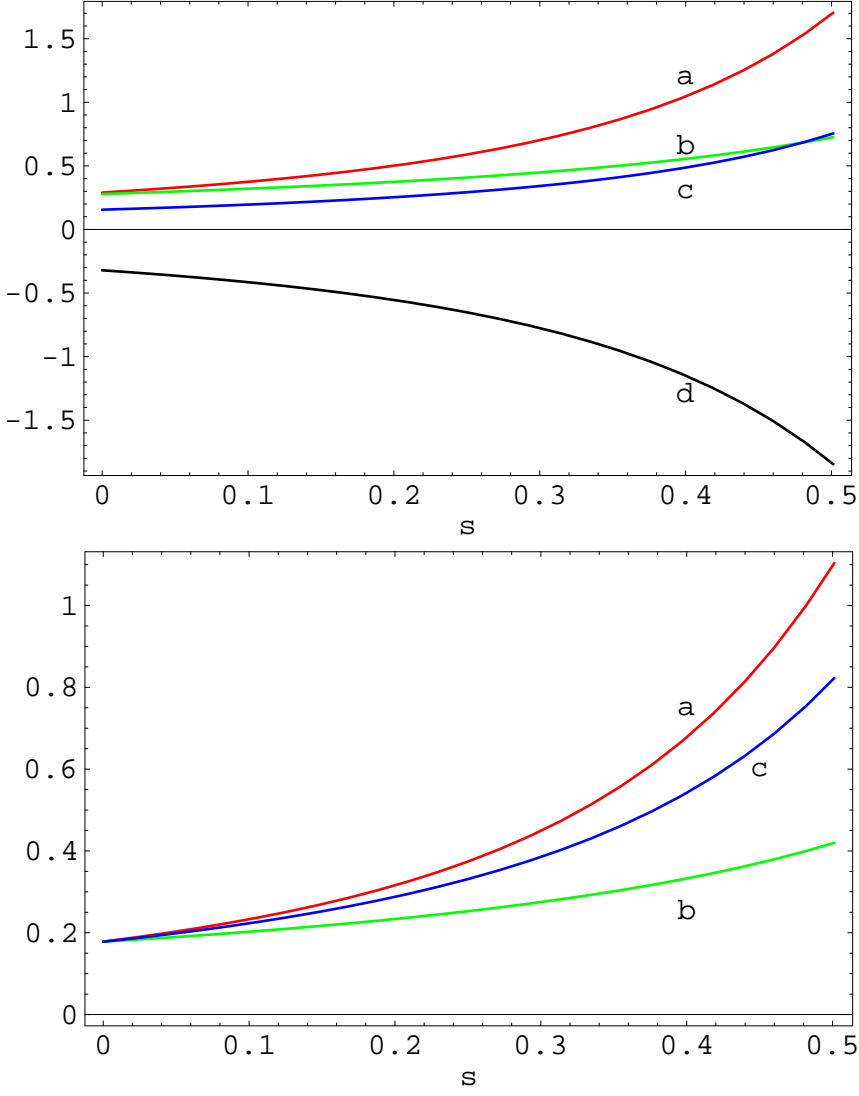}
\vspace*{-3cm}
\caption{
Form factors for the $B_c\to D_d^\ast$ transition. 
Upper panel:  
$(a)\, V$, $(b)\, A_0$, $(c)\, A_+$, $(d)\, A_-$.
Lower panel: $(a)\, g$, $(b)\, a_0$, $(c)\, a_+$.
}
\label{fig:bc_dv}
\end{figure}

\begin{figure}
\vspace*{19.5cm}
\includegraphics{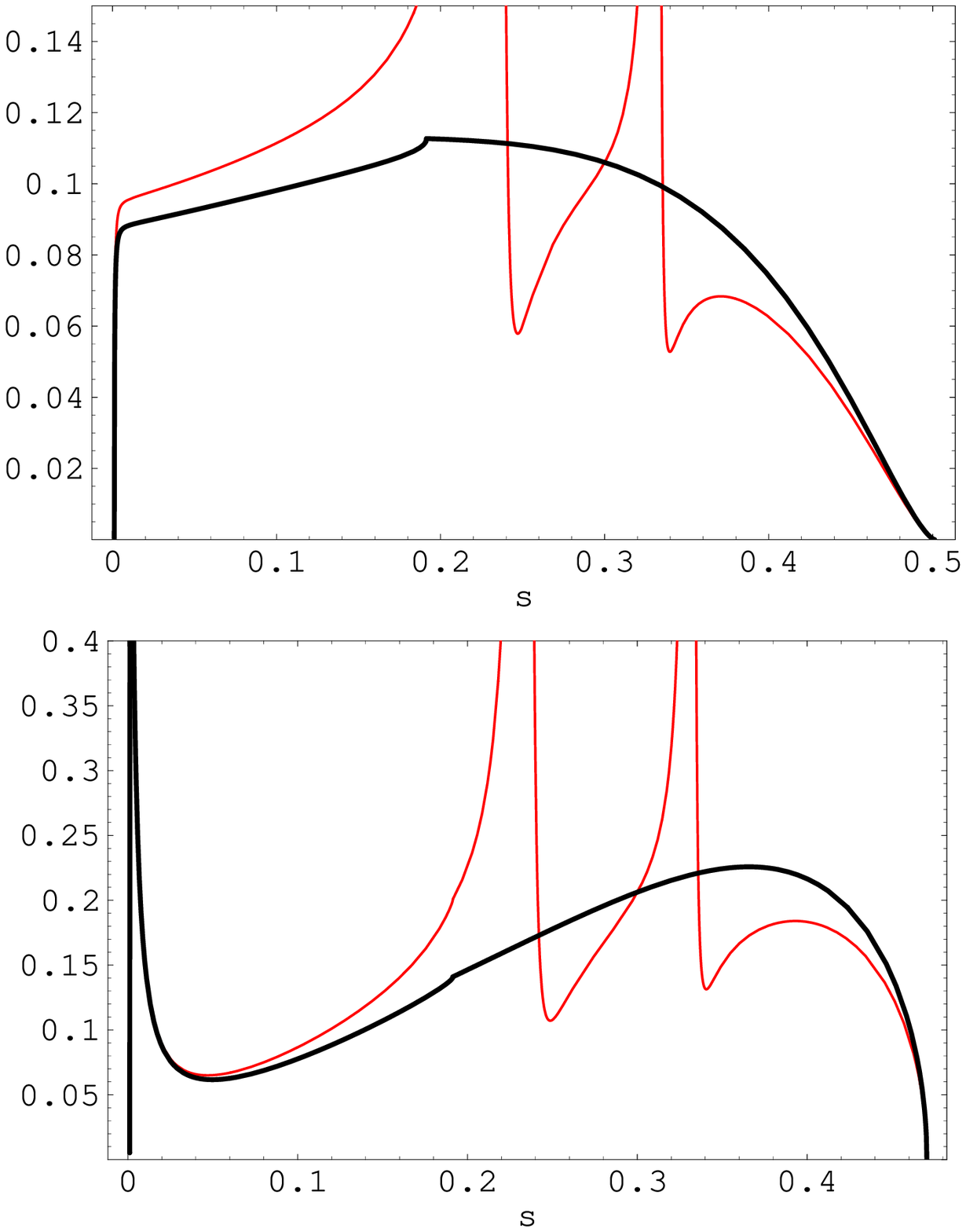}
\vspace*{-3cm}
\caption{
Normalized differential distributions 
$10^7\,\Gamma^{-1}_{\rm tot} d{\Gamma}/ds$ 
for $B_c\to D\, \mu^+\mu^-$ (upper panel) and 
$B_c\to D^\ast\,\mu^+\mu^-$ (lower panel).}
\label{fig:bc_ll}
\end{figure}

\begin{figure}
\vspace*{19.5cm}
\includegraphics{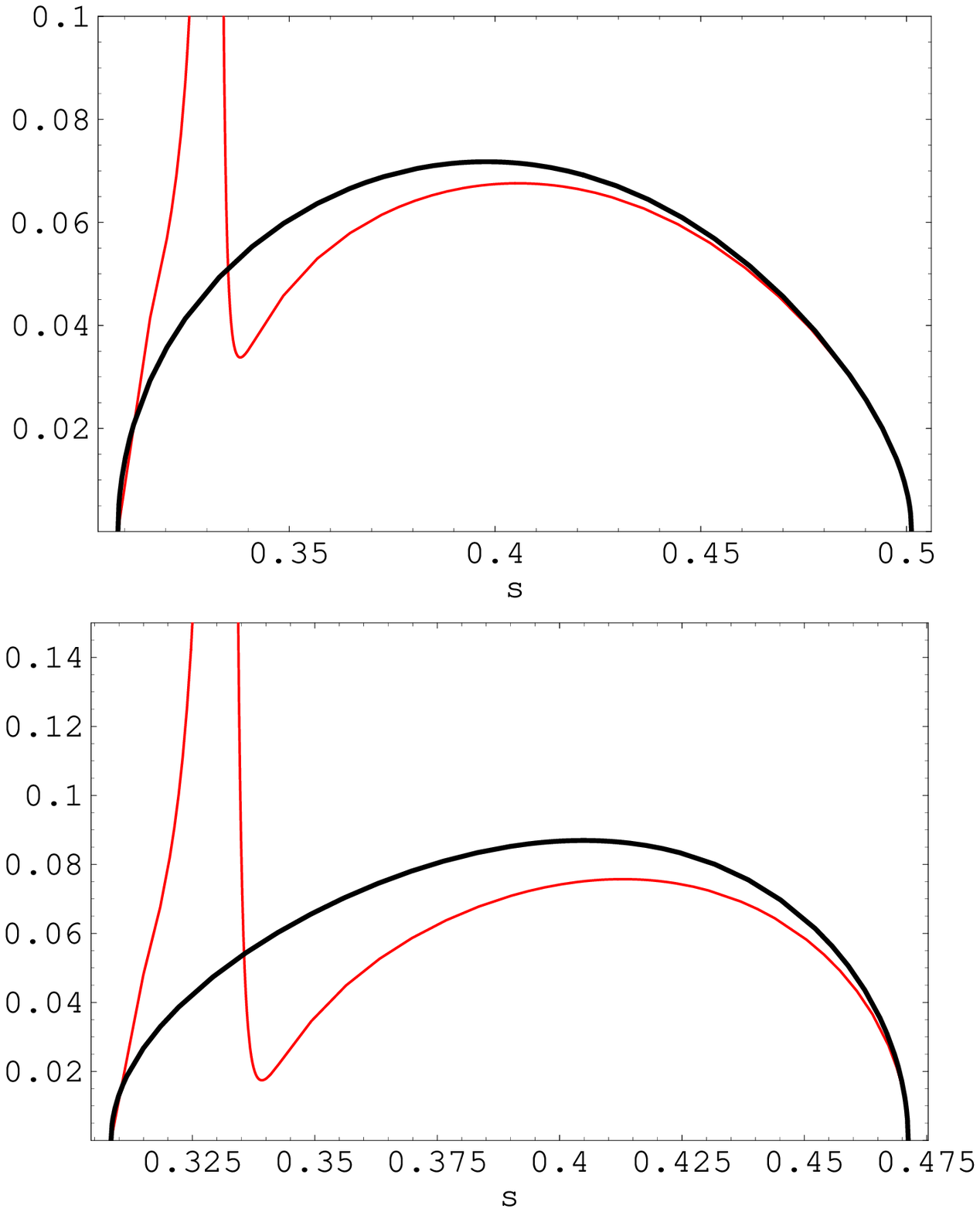}
\vspace*{-3cm}
\caption{
Normalized differential distributions 
$10^7\,\Gamma^{-1}_{\rm tot} d{\Gamma}/ds$ 
for $B_c\to D\, \tau^+\tau^-$ (upper panel) and 
$B_c\to D^\ast\,\tau^+\tau^-$ (lower panel).
}
\label{fig:bc_tau}
\end{figure}

\begin{figure}
\vspace*{19.5cm}
\includegraphics{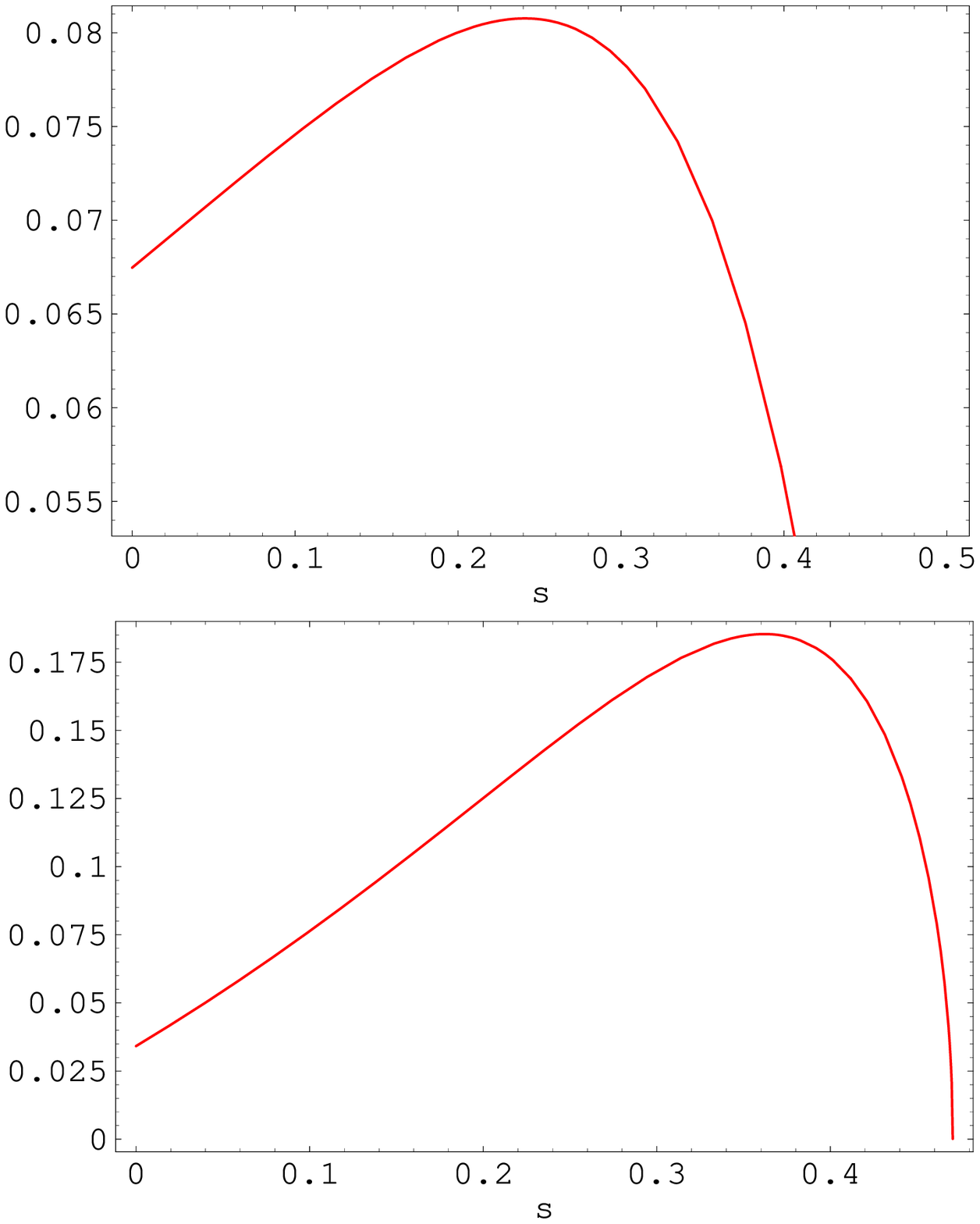}
\vspace*{-3cm}
\caption{
Normalized differential distributions 
$10^7\,\Gamma^{-1}_{\rm tot} d{\Gamma}/ds$ 
for $B_c\to D\,\bar\nu\nu$ (upper panel) and 
$B_c\to D^\ast\,\bar\nu\nu$ (lower panel).
}
\label{fig:bc_nn}
\end{figure}

Finally, we plot the dependence of the normalized differential 
distributions $s=q^2/m^2_{B_c}$ in Figs.~\ref{fig:bc_ll}-\ref{fig:bc_nn} 
for the decay $B_c\to D(D^\ast) \bar ll (\bar \nu \nu)$.
In the numerical analysis we use the input
parameters: $m_{B_c}=6.4$ GeV, $\tau_{B_c}=0.46$ ps and
$|V^\dagger_{\rm td}V_{\rm tb}|=0.008$.
The $B_c\to D_d( D_d^\ast)$-transition form factors are plotted
in Figs.~\ref{fig:bc_d},\ref{fig:bc_dv} and the normalized differential
distributions for $B_c\to D(D^\ast)\mu^+\mu^-$, 
$B_c\to D(D^\ast)\tau^+\tau^-$ and $B_c\to D(D^\ast)\bar\nu\nu$ 
are shown in  Figs.~\ref{fig:bc_ll},\ref{fig:bc_tau},\ref{fig:bc_nn}, 
respectively.
The results for the branching ratios are also given in 
Table~\ref{tab:branching}.
They are to be compared with the  results of calculations 
performed in  \cite{Geng:2001vy}
where the light front and constituent quark models were employed.

\section{Acknowledgments}

This work was completed while M.A.I. visited the Universities 
of Mainz and T\"{u}bingen. M.A.I. appreciates the partial support
by Plester Foundation, the  DFG grants GRK683 and 436 RUS17/47/02, 
the Russian Fund of Basic Research grant No. 01-02-17200 and the 
Heisenberg-Landau Program. A.F., Th.G. and V.E.L. thank the DFG  
grants FA67/25-1 and GRK683 for support.

\newpage

\appendix{\large\bf Appendix A: Wilson Coefficients} 
\vspace*{1cm}

In this paper we use the Wilson-coefficients $C_i$ calculated in the 
naive dimensional regularization (NDR) scheme 
in the leading logarithmic approximation \cite{Buras:1995}:
 
\begin{eqnarray*} 
C_j(\mu)    & = & \sum_{i=1}^8 k_{ji} \eta^{a_i}
  \qquad (j=1,...6) \vspace{0.2cm} 
\\
C_7(\mu) & = & \eta^\frac{16}{23} C_7(M_W) + \frac{8}{3}
   \left(\eta^\frac{14}{23} - \eta^\frac{16}{23}\right) C_8(M_W) +
   \sum_{i=1}^8 h_i \eta^{a_i},
\end{eqnarray*}

with

$$
\eta  =  \frac{\alpha(M_W)}{\alpha(\mu)}, \hspace{1cm}
C_7(M_W)  =  - \frac{1}{2} A(x_t), \hspace{1cm}
C_8(M_W)  =  - \frac{1}{2} F(x_t),
$$
where $x_t = m_t^2/M_W^2$ and $A(x)$ and $F(x)$ are defined below.
The numbers $a_i$, $k_{ji}$ and $h_i$  are given in 
Table~\ref{tab:wilson}.

The coefficient of $Q_{10}$ is given by
$$
C_{10}(M_W) = - \frac{Y(x_t)}{\sin^2\Theta_W}
$$
with $Y(x)$ given below. Since $Q_{10}$ is not subject to renormalization
under QCD, its coefficient does not depend on $\mu\approx {\cal O}(m_b)$. 
The only renormalization scale dependence 
enters through the definition of the top quark mass. 

Finally, including leading as well as next-to-leading logarithms, 
one finds
$$
C_9(\mu)  = 
P_0 + \frac{Y(x_t)}{\sin^2\Theta_W} -4 Z(x_t) + P_E E(x_t)
$$
with
\begin{eqnarray*}
P_0 & = & \frac{\pi}{\alpha_s(M_W)} (-0.1875+ \sum_{i=1}^8 p_i
\eta^{a_i+1}) 
 + 1.2468 +  \sum_{i=1}^8 \eta^{a_i} \lbrack
r_i+s_i \eta \rbrack \nonumber\\
P_E & = & 0.1405 +\sum_{i=1}^8 q_i\eta^{a_i+1} \nonumber\\
Y(x) & = & C(x) - B(x), \hspace{1cm} Z(x) = C(x) + \frac{1}{4} D(x).
\end{eqnarray*}
Here
\begin{eqnarray*}
A(x) & = & \frac{x(8x^2+5x-7)}{12(x-1)^3} + \frac{x^2(2-3x)}{2(x-1)^4} \ln
x, \\
B(x) & = & \frac{x}{4(1-x)} + \frac{x}{4(x-1)^2} \ln x,\\
C(x) & = & \frac{x(x-6)}{8(x-1)} + \frac{x(3x+2)}{8(x-1)^2} \ln x, \\
D(x) & = & \frac{-19 x^3 + 25 x^2}{36 (x-1)^3}
+ \frac{x^2 (5 x^2 - 2 x - 6)}{18 (x-1)^4} \ln x- \frac{4}{9} \ln x,\\
E(x) & = & \frac{x (18 -11
x - x^2)}{12 (1-x)^3} + \frac{x^2 (15 - 16 x + 4 x^2)}{6 (1-x)^4} \ln
x-\frac{2}{3} \ln x, \\
F(x) & = & \frac{x(x^2-5x-2)}{4(x-1)^3} + \frac{3x^2}{2(x-1)^4} \ln x.
\end{eqnarray*}

The coefficients $p_i$, $r_i$, $s_i$, and $q_i$ are given in 
Table~\ref{tab:wilson}.

\vspace{1cm}
\noindent
\appendix{\large\bf Appendix B: The functions in Eqs.~(\ref{check_1}) 
and (\ref{check_2}).}
\vspace{1cm}

We list here a set of the functions appearing in 
Eqs.~(\ref{check_1}) and (\ref{check_2}) 
from  \cite{Ali:1999mm} and \cite{Geng:2001vy}.

\begin{eqnarray*}
\phi_H &=&\biggl( 1-r_H\biggr) ^{2}-2s\biggl( 1+r_H\biggr) +s^{2},
     \\
&&\\
D_\nu(x) &=& \frac{x}{8}\biggl(\frac{2+x}{x-1}+\frac{3x-6}{(x-1)^2}
\,\ln x\biggr)\\
&&\\
\alpha_1 &=& (1-\sqrt{r_H}) ^{2} \biggl| A_{0}\biggr| ^{2} +
           \frac{\phi_H
           }{(1+\sqrt{r_H})^{2}}\biggl| V\biggr| ^{2}
       \\
    \beta_1 &=& \frac{(1-\sqrt{r_H})^2}{4r_H} \biggl| A_{0}\biggr|^2 
                       -\frac{s}{(1+\sqrt{r_H})^2}\biggl| V\biggr| ^2
           +\frac{\phi _H \biggl| A_+\biggr| ^2}{4r_H(1+\sqrt{r_H})^2}
           \nonumber \\
           &-& \,\frac{1}{2}\biggl( \frac{1-s}{r_H}-1\biggr)
           \frac{1-\sqrt{r_H}}{1+\sqrt{r_H}}
          {\rm Re}(A_{0}A_{+}^\dagger)\,,\\
&&\\
    \alpha_2 &=&
\biggl|C_{9}^{\rm eff}\,F_{+}
+\frac{2\,{\hat m}_b\,C_{7}^{\rm eff}\,F_{T}}{1+\sqrt{r_H}}\biggr|^{2}
       +|C_{10}F_{+}|^{2} ~,
  \\
    \beta_2 &=& |C_{10}|^{2}\biggl[ \biggl(
       1+r_H-{s\over 2}\biggr) |F_{+}|^{2}+\biggl( 1-r_H\biggr)
       {\rm Re}(F_{+}F_{-}^\dagger)+\frac{1}{2}s|F_{-}|^{2}\biggr]\,,
\\
 && \\
  \alpha_3 &=& (1-\sqrt{r_H})^2\biggl[ \biggl| C_9^{\rm eff}A_0
     +\frac{2{\hat m}_b\,C_7^{\rm eff}\,(1+\sqrt{r_H})\,a_0}{s}\biggr|^2
     +\biggl| C_{10}A_0\biggr| ^2\biggr]
     \\
     &+&\frac{\phi _H}{(1+\sqrt{r_H})^2}\biggl[ \biggl| C_9^{\rm eff}V
     +\frac{2{\hat m}_b\,C_7^{\rm eff}\,(1+\sqrt{r_H})\,g}{s}\biggr| ^2
     +\biggl| C_{10}V\biggr|^2\biggr] \,,
      \\
  \beta_3 &=& \frac{(1-\sqrt{r_H})^2}{4r_H}\biggl[ 
     \biggl| C_9^{\rm eff}A_0
     +\frac{2{\hat m}_b\,C_7^{\rm eff}(1+\sqrt{r_H})\,a_0}{s}\biggr|^2
     +\biggl| C_{10}A_0\biggr|^2\biggr]
     \nonumber \\
     &-&\frac{s}{(1+\sqrt{r_H})^2}\biggl[ \biggl| C_9^{\rm eff}V
     +\frac{2{\hat m}_b\,C_7^{\rm eff}(1+\sqrt{r_H})\,g}{s}\biggr|^2
     +\biggl| C_{10}V\biggr|^2\biggr]
     \nonumber \\
     &+&\frac{\phi _H}{4r_H(1+\sqrt{r_H})^2}
     \biggl[\biggl|C_9^{\rm eff}A_+ 
     +\,\frac{2{\hat m}_b\,C_7^{\rm eff}\,(1+\sqrt{r_H})\,a_+}{s}
     \biggr|^2 +\biggl| C_{10}A_+\biggr| ^2\biggr]
     \nonumber \\
     &-&\,\frac{1}{2}\biggl(\frac{1-s}{r_H}-1\biggr)
          \frac{1-\sqrt{r_H}}{1+\sqrt{r_H}}
     {\rm Re}\biggl\{ \biggl[C_9^{\rm eff}A_0 
      +\,\frac{2{\hat m}_b\,C_7^{\rm eff}\,
    (1+\sqrt{r_H})\,a_0}{s}\biggr]\\
     &\times&\biggl[C_9^{\rm eff}A_+ 
     +\,\frac{2{\hat m}_b\,C_7^{\rm eff}\,(1+\sqrt{r_H})\,a_+}{s}\biggr]
     +|C_{10}|^2\,{\rm Re}(A_0A_+^\dagger)\biggr\}
     \,,\\
&&\\
  \delta &=& \frac{|C_{10}|^2}{2(1+\sqrt{r_H})^2}
\biggl\{ -2\phi_H\,|V|^2-3(1-r_H)^2\,|A_0|^2\\
&+&\frac{\phi _H}{4r_H}\,\biggl[2(1+r_H)-s\biggr]\,|A_+|^2 \\
     &+& \frac{\phi _Hs}{4r_H}|A_-|^2+\frac{\phi _H(1-r_H)}{2r_H}
{\rm Re}
\biggl(-\, A_0A_+^\dagger-\,A_0A_-^\dagger +\,A_+A_-^\dagger\biggr) 
\biggr\}\,,
\end{eqnarray*}
where ${\hat m}_b=m_b/m_{B}$.

\newpage


\begin{table}
\caption{Central values of the input parameters and the corresponding
values of the Wilson coefficients used in the numerical calculations.}  
        \begin{center}
        \begin{tabular}{|c|c||l|r|}
\hline 
 $m_W$                   & 80.41   GeV      & $C_1$    & -0.248   \\   
 $m_Z$                   & 91.1867 GeV      & $C_2$    &  1.107   \\
 $\sin^2 \theta_W $      & 0.2233           & $C_3$    &  0.011   \\
 $m_c$                   & 1.4     GeV      & $C_4$    & -0.026   \\
 $m_t$                   & 173.8   GeV      & $C_5$    &  0.007   \\
 $m_{b,\rm  pole}$       & 4.8     GeV      & $C_6$    & -0.031   \\
 $\mu$                   & $m_{b,\rm pole}$ & $C_7^{\rm eff}$ & -0.313   \\
 $\Lambda_{QCD}$         & 0.220   GeV      & $C_9$    &  4.344   \\
 $\alpha^{-1}$           & 129              & $C_{10}$ & -4.669   \\
 $\alpha_s (m_Z) $       & 0.119            & $C_0$    &  0.362   \\
 $|V^\dagger_{ts} V_{tb}|$  & 0.0385           &          &          \\
 $|V^\dagger_{td} V_{tb}|$  & 0.008            &          &          \\
 $|V^\dagger_{ts} V_{tb}|/|V_{cb}| $ & 1       &          &          \\
\hline 
\end{tabular}
\label{tab:input}
\end{center}
\end{table}

\begin{table}
\begin{center}
\caption{Bilinear combinations of the helicity amplitudes
that enter in the four-fold decay distribution Eq.~( \ref{distr4} ).}
\vspace*{.25cm}
\def\arraystretch{1}
\begin{tabular}{|l|l|l|}
\hline
Definition & Property & Title \\
\hline
$ H_{U}^{ij}={\rm Re}\left(H_{+}^i H_{+}^{\dagger\,j}\right)
            +{\rm Re}\left(H_{-}^i H_{-}^{\dagger\,j}\right) $ &
$ H_{U}^{ij}=  H_{U}^{ji}$ & {\bf U}npolarized-\\[-2.5mm]
& & transverse \\[-1.5mm]
$ H_{IU}^{ij}={\rm Im}\left(H_{+}^i H_{+}^{\dagger\,j}\right)
             +{\rm Im}\left(H_{-}^i H_{-}^{\dagger\,j}\right) $ &
$ H_{IU}^{ij}=-H_{IU}^{ji}$ &  \\
\hline
$ H_{P}^{ij}={\rm Re}\left(H_{+}^i H_{+}^{\dagger\,j}\right)
            -{\rm Re}\left(H_{-}^i H_{-}^{\dagger\,j}\right) $ &
$ H_{P}^{ij}=  H_{P}^{ji}$ & {\bf P}arity-odd \\
$ H_{IP}^{ij}={\rm Im}\left(H_{+}^i H_{+}^{\dagger\,j}\right)
             -{\rm Im}\left(H_{-}^i H_{-}^{\dagger\,j}\right) $ &
$ H_{IP}^{ij}=-H_{IP}^{ji}$ &  \\
\hline
$ H_{T}^{ij}={\rm Re}\left(H_{+}^i H_{-}^{\dagger\,j}\right)$ & & 
{\bf T}ransverse-\\[-2.5mm]
& & interference \\[-1.5mm]
$ H_{IT}^{ij}={\rm Im}\left(H_{+}^i H_{-}^{\dagger\,j}\right)$ & & \\
\hline
$ H_{L}^{ij}={\rm Re}\left(H_{0}^i H_{0}^{\dagger\,j}\right) $ &
$ H_{L}^{ij}= H_{L}^{ji}$ & {\bf L}ongitudinal \\
$ H_{IL}^{ij}={\rm Im}\left(H_{0}^i H_{0}^{\dagger\,j}\right)$ &
$ H_{IL}^{ij}= -H_{IL}^{ji}$ & \\
\hline
$ H_{S}^{ij}=3\,{\rm Re}\left(H_{t}^i H_{t}^{\dagger\,j}\right) $ &
$ H_{S}^{ij}= H_{S}^{ji}$ & {\bf S}calar \\
$ H_{IS}^{ij}=3\,{\rm Im}\left(H_{t}^i H_{t}^{\dagger\,j}\right)$ &
$ H_{IS}^{ij}= -H_{IS}^{ji}$ & \\
\hline
$ H_{SL}^{ij}={\rm Re}\left(H_{t}^i H_{0}^{\dagger\,j}\right) $ & & 
{\bf S}calar-\\[-2.5mm]
& &{\bf L}ongitudinal-\\[-1.5mm]
& & interference \\[-1.5mm]
$ H_{ISL}^{ij}={\rm Im}\left(H_{t}^i H_{0}^{\dagger\,j}\right)$ & & \\
\hline
$ H_{I}^{ij}=\frac{1}{2}\,
 \left[{\rm Re}\left(H_{+}^i H_{0}^{\dagger\,j}\right)
      +{\rm Re}\left(H_{-}^i H_{0}^{\dagger\,j}\right)\right]$ & &
  transverse-\\[-2.5mm]
& &longitudinal-\\[-1.5mm]
& & {\bf I}nterference  \\[-1.5mm]
$ H_{II}^{ij}=\frac{1}{2}\,
\left[{\rm Im}\left(H_{+}^i H_{0}^{\dagger\,j}\right)
     +{\rm Im}\left(H_{-}^i H_{0}^{\dagger\,j}\right)\right]$ 
&  &  \\
\hline
$ H_{A}^{ij}=\frac{1}{2}\,
\left[{\rm Re}\left(H_{+}^i H_{0}^{\dagger\,j}\right)
     -{\rm Re}\left(H_{-}^i H_{0}^{\dagger\,j}\right)\right]$ 
&  & parity-\\[-2.5mm]
& &{\bf A}symmetric\\[-1.5mm]
$ H_{IA}^{ij}=\frac{1}{2}\,
\left[{\rm Im}\left(H_{+}^i H_{0}^{\dagger\,j}\right)
     -{\rm Im}\left(H_{-}^i H_{0}^{\dagger\,j}\right)\right]$ 
&  &  \\
\hline
$ H_{ST}^{ij}=\frac{1}{2}\,
\left[{\rm Re}\left(H_{+}^i H_{t}^{\dagger\,j}\right)
     +{\rm Re}\left(H_{-}^i H_{t}^{\dagger\,j}\right)\right]$ 
& & {\bf S}calar-\\[-2.5mm]
& &{\bf T}ransverse-\\[-1.5mm]
& & interference  \\[-1.5mm]
$ H_{IST}^{ij}=\frac{1}{2}\,
\left[{\rm Im}\left(H_{+}^i H_{t}^{\dagger\,j}\right)
     +{\rm Im}\left(H_{-}^i H_{t}^{\dagger\,j}\right)\right]$ 
&  &  \\
\hline
$ H_{SA}^{ij}=\frac{1}{2}\,
\left[{\rm Re}\left(H_{+}^i H_{t}^{\dagger\,j}\right)
     -{\rm Re}\left(H_{-}^i H_{t}^{\dagger\,j}\right)\right]$ 
& & {\bf S}calar-\\[-2.5mm]
& &{\bf A}symmetric-\\[-1.5mm]
& & interference  \\[-1.5mm]
$ H_{ISA}^{ij}=\frac{1}{2}\,
\left[{\rm Im}\left(H_{+}^i H_{t}^{\dagger\,j}\right)
     -{\rm Im}\left(H_{-}^i H_{t}^{\dagger\,j}\right)\right]$ 
&  &  \\
\hline
\end{tabular}
\label{tab:helicity}
\end{center}
\end{table}

\newpage 

\begin{table}[ht]
\caption{
Leptonic decay constants $f_H$ (MeV) used in the least-square fit.
The values are taken either from PDG~\cite{Hagiwara:in} or
from the Lattice~\cite{Ryan}: quenched (upper line)
and  unquenched (lower line). }
\begin{center}
\begin{tabular}{|l|l|l|}
\hline 
Meson & This model & Expt/Lattice \\
\hline 
$\pi^+$      & 131        & $130.7\pm 0.1\pm 0.36$    \\
\hline
$K^+$        & 161        & $159.8\pm 1.4\pm 0.44$        \\
\hline
$D^+$        & 211        & 203$\pm$ 14 \\
             &            & 226$\pm$ 15 \\
\hline 
$D^+_s$      & 222        &  230$\pm$ 14 \\
             &            &  250$\pm$ 30 \\
\hline   
$B^+$        & 180        &  173$\pm$ 23 \\
             &            &  198$\pm$ 30 \\
\hline   
$B^0_s$      & 196        &  200$\pm$ 20 \\
             &            &  230$\pm$ 30 \\
\hline
$B^+_c$      & 398        &         \\
\hline 
\end{tabular}
\label{tab:fit}
\end{center}
\end{table}

\begin{table}[ht]
\caption{
Semileptonic decay branching ratios.
}
\begin{center}
\begin{tabular}{|l|l|l|}
\hline 
Meson & This model & Expt. \\
\hline 
$\pi^+\to \pi^0 l^+ \nu$         & $1.03\cdot 10^{-8}$         & 
                           $(1.025\pm 0.034)\cdot 10^{-8}$    \\
\hline
$K^+  \to \pi^0 l^+ \nu$         & $4.62\cdot 10^{-2}$         & 
                           $(4.82\pm 0.06)\cdot 10^{-2}$    \\
\hline
$B^+\to \bar D^0 l^+ \nu$         & $2.40\cdot 10^{-2}$         & 
                           $(2.15\pm 0.22)\cdot 10^{-2}$    \\
\hline
$B^+\to \bar D^{\ast\,0} l^+ \nu$ & $5.60\cdot 10^{-2}$         & 
                           $(5.3\pm 0.8)\cdot 10^{-2}$    \\
\hline
$B_c^+\to  D^0 l^+ \nu$         & $2.05\cdot 10^{-5}$         & \\ 
\hline
$B_c^+\to  D^{\ast\,0} l^+ \nu$ & $3.60 \cdot 10^{-5}$        & \\
\hline 
\end{tabular}
\label{tab:sem}
\end{center}
\end{table}

\begin{table}[ht]
\caption{Parameter values for the approximated form factors 
$F(s)=F(0)/(1-a s + b s^2)$\,
$(s=q^2/m_B^2)$ in the  decays $B\to K \bar l l$.}
\begin{center}
\begin{tabular}{|l|lll|}
\hline 
 & $F_+$ & $F_-$ & $F_T$ \\ 
\hline
$F(0)$ & 0.357  & -0.275 & 0.337 \\
$a$    & 1.011  &  1.050 & 1.031 \\
$b$    & 0.042  &  0.067 & 0.051 \\
\hline 
\end{tabular}
\label{tab:ffbk}
\end{center}
\end{table}

\begin{table}[ht]
\caption{Parameter values for the approximated form factors 
 $F(s)=F(0)/(1-a s + b s^2)$ \,
$(s=q^2/m_B^2)$ in the  decays $B_c\to D(D^\ast) \bar l l$.}
\begin{center}
\begin{tabular}{|l|lll|llll|lll|}
\hline 
 & $F_+$ & $F_-$ & $F_T$ & $A_0$ & $A_+$ & $A_-$ & $V$ & 
   $a_0$ & $a_+$ & $g$ \\ 
\hline
$F(0)$ & 0.186  & -0.190 & 0.275 & 0.279  & 0.156 & 
        -0.321  &  0.290 & 0.178 & 0.178  & 0.179 \\
$a$    & 2.48   &  2.44  & 2.40  & 1.30   & 2.16  & 
         2.41   &  2.40  & 1.21  & 2.14   & 2.51   \\
$b$    & 1.62   &  1.54  & 1.49  & 0.149  & 1.15  & 
         1.51   &  1.49  & 0.125 & 1.14   & 1.67   \\
\hline 
\end{tabular}
\label{tab:ffbcd}
\end{center}
\end{table}

\begin{table}[ht]
\caption{Decay branching ratios without(with) long distance
contributions.}
\begin{center}
\begin{tabular}{|c|c|c|c|}
\hline 
Ref. &
 ${\rm Br}(B\to K\,\mu^+\mu^-)$ & ${\rm Br}(B\to K\,\tau^+\tau^-)$ 
 & ${\rm Br}(B\to K\,\bar\nu\nu)$ \\
\hline\hline
\cite{Ali:1999mm} & $0.57\cdot 10^{-6}$ &$1.3\cdot 10^{-7}$ & \\
\hline
\cite{Ali:2001jg} & 
$(0.35\pm 0.12)\cdot 10^{-6}$ &  & \\ 
\hline
\cite{Melikhov:1997wp} &$0.44\cdot 10^{-6}$ & $1.0\cdot 10^{-7}$ 
& $5.6\cdot 10^{-6}$  \\ 
\hline
\cite{Geng:1996} & $0.5\cdot 10^{-6}$ & $1.3\cdot 10^{-7}$ & \\
\hline
our & $0.55\,(0.51)\cdot 10^{-6}$ & $1.01\,(0.87)\cdot 10^{-7}$ 
& $4.19\cdot 10^{-6}$ \\
\hline 
\end{tabular}
\end{center}

\begin{center}
\begin{tabular}{|c|c|c|}
\hline 
      & our & \cite{Geng:2001vy} \\
\hline\hline
${\rm Br}(B_c\to D_d\,\mu^+\mu^-)$ &        $0.44\,(0.38)\cdot 10^{-8}$ &  
                                            $0.41\,(0.33)\cdot 10^{-8}$ \\
${\rm Br}(B_c\to D_d^\ast\,\mu^+\mu^-)$ &   $0.71\,(0.58)\cdot 10^{-8}$ &  
                                            $1.01\,(0.78)\cdot 10^{-8}$ \\
${\rm Br}(B_c\to D_s\,\mu^+\mu^-)$ &        $0.97\,(0.86)\cdot 10^{-7}$ &  
                                            $1.36\,(1.12)\cdot 10^{-7}$ \\
${\rm Br}(B_c\to D_s^\ast\,\mu^+\mu^-)$ &   $1.76\,(1.41)\cdot 10^{-7}$ &  
                                            $4.09\,(3.14)\cdot 10^{-7}$ \\
\hline
${\rm Br}(B_c\to D_d\,\tau^+\tau^-)$ &      $0.11\,(0.09)\cdot 10^{-8}$ &  
                                            $0.13\,(0.11)\cdot 10^{-8}$ \\
${\rm Br}(B_c\to D_d^\ast\,\tau^+\tau^-)$ & $0.11\,(0.08)\cdot 10^{-8}$ &  
                                            $0.18\,(0.13)\cdot 10^{-8}$ \\
${\rm Br}(B_c\to D_s\,\tau^+\tau^-)$ &      $0.22\,(0.18)\cdot 10^{-7}$ &  
                                            $0.34\,(0.27)\cdot 10^{-7}$ \\
${\rm Br}(B_c\to D_s^\ast\,\tau^+\tau^-)$ & $0.22\,(0.15)\cdot 10^{-7}$ &  
                                            $0.51\,(0.34)\cdot 10^{-7}$ \\
\hline\hline
${\rm Br}(B_c\to D_d\,\bar\nu\nu)$       & $3.28\cdot 10^{-8}$ &\\
${\rm Br}(B_c\to D_d^\ast\,\bar\nu\nu)$  & $5.78\cdot 10^{-8}$ &\\
${\rm Br}(B_c\to D_s\,\bar\nu\nu)$       & $0.73\cdot 10^{-6}$ &\\
${\rm Br}(B_c\to D_s^\ast\,\bar\nu\nu)$  & $1.42\cdot 10^{-6}$ &\\
\hline 
\end{tabular}
\label{tab:branching}
\end{center}
\end{table}

\begin{table}[ht]
\caption{
Values of  parameters in the formulae for the Wilson
coefficients.}
\begin{center}
\begin{tabular}{|r|r|r|r|r|r|r|r|r|}
\hline\hline 
$i$ & 1 & 2 & 3 & 4 & 5 & 6 & 7 & 8 \\
\hline\hline
$a_i $&$ \frac{14}{23} $&$ \frac{16}{23} $&$ \frac{6}{23} $&$
-\frac{12}{23} $&$
0.4086 $&$ -0.4230 $&$ -0.8994 $&$ 0.1456 $\\
$k_{1i} $&$ 0 $&$ 0 $&$ \frac{1}{2} $&$ - \frac{1}{2} $&$
0 $&$ 0 $&$ 0 $&$ 0 $\\
$k_{2i} $&$ 0 $&$ 0 $&$ \frac{1}{2} $&$  \frac{1}{2} $&$
0 $&$ 0 $&$ 0 $&$ 0 $\\
$k_{3i} $&$ 0 $&$ 0 $&$ - \frac{1}{14} $&$  \frac{1}{6} $&$
0.0510 $&$ - 0.1403 $&$ - 0.0113 $&$ 0.0054 $\\
$k_{4i} $&$ 0 $&$ 0 $&$ - \frac{1}{14} $&$  - \frac{1}{6} $&$
0.0984 $&$ 0.1214 $&$ 0.0156 $&$ 0.0026 $\\
$k_{5i} $&$ 0 $&$ 0 $&$ 0 $&$  0 $&$
- 0.0397 $&$ 0.0117 $&$ - 0.0025 $&$ 0.0304 $\\
$k_{6i} $&$ 0 $&$ 0 $&$ 0 $&$  0 $&$
0.0335 $&$ 0.0239 $&$ - 0.0462 $&$ -0.0112 $\\
\hline
$h_i $&$ 2.2996 $&$ - 1.0880 $&$ - \frac{3}{7} $&$ -
\frac{1}{14} $&$ -0.6494 $&$ -0.0380 $&$ -0.0185 $&$ -0.0057 $\\
$\bar h_i $&$ 0.8623 $&$ 0 $&$ 0 $&$ 0
 $&$ -0.9135 $&$ 0.0873 $&$ -0.0571 $&$ 0.0209 $\\
\hline
$p_i $&$ 0 $&$ 0 $&$ -\frac{80}{203} $&$  \frac{8}{33} $&$
0.0433 $&$  0.1384 $&$ 0.1648 $&$ - 0.0073 $\\
$r_{i} $&$ 0 $&$ 0 $&$ 0.8966 $&$ - 0.1960 $&$
- 0.2011 $&$ 0.1328 $&$ - 0.0292 $&$ - 0.1858 $\\
$s_i $&$ 0 $&$ 0 $&$ - 0.2009 $&$  -0.3579 $&$
0.0490 $&$ - 0.3616 $&$ -0.3554 $&$ 0.0072 $\\
$q_i $&$ 0 $&$ 0 $&$ 0 $&$  0 $&$
0.0318 $&$ 0.0918 $&$ - 0.2700 $&$ 0.0059 $ \\
\hline 
\end{tabular}
\label{tab:wilson}
\end{center}
\end{table}

\clearpage

\newpage 


\newpage 

\newpage

\newpage 

\newpage 

\newpage 

\newpage 

\newpage 

\newpage 

\end{document}